\documentclass[cernpreprint, atlasdraft=false, UKenglish, texmf, orcidlogo]{atlasdoc}
 
\usepackage[subcaption]{atlaspackage}
\usepackage{booktabs, multirow}
\usepackage{float}
\usepackage{atlasbiblatex}
 
\usepackage[jetetmiss=true]{atlasphysics}
 
\graphicspath{{logos/}{figures/}}
 
\usepackage{ANA-JETM-2023-02-PAPER-defs}
\usepackage{siunitx}

\AtlasTitle{Simultaneous energy and mass calibration of large-radius jets with the ATLAS detector using a deep neural network}
 
\AtlasAbstract{
The energy and mass measurements of jets are crucial tasks for the Large Hadron Collider experiments. This paper presents a new calibration method to simultaneously calibrate these quantities for large-radius jets measured with the ATLAS detector using a deep neural network (DNN).
To address the specificities of the calibration problem, special loss functions and training procedures are employed, and a complex network architecture, which includes feature annotation
and residual connection layers, is used. The DNN-based calibration is compared to the standard numerical approach in an extensive series of tests. The DNN approach is found to perform significantly better in almost all of the tests and over most of the relevant kinematic phase space. In particular, it consistently improves the energy and mass resolutions, with a 30\% better energy resolution obtained for transverse momenta $p_{\text{T}}>500$ GeV.
}
 
\author{The ATLAS Collaboration}
 
\AtlasRefCode{JETM-2023-02}
 
\PreprintIdNumber{CERN-EP-2023-250}

\arXivId{2311.08885}

\AtlasJournalRef{Mach. Learn. Sci. Technol. 5 035051 (2024)}
\AtlasDOI{10.1088/2632-2153/ad611e}

\hypersetup{pdftitle={ATLAS document},pdfauthor={The ATLAS Collaboration}}
 
\begin{document}
 
\maketitle
 
\tableofcontents
 
\newpage

 
\section{Introduction}
Jets -- the collimated sprays of hadrons resulting
from a high momentum quark or gluon emission -- are abundantly produced
during the operation of the Large Hadron Collider (LHC). They are relevant to
almost all physics analyses of the collisions recorded by the LHC
experiments, either because they are part of the final state of the studied
signal or because they generate background for the process
of interest.
To achieve high-quality physics results, it is crucial to  precisely
measure the angular position and energy of the jets. This task is made difficult
by multiple effects including fluctuations in the parton showers or in the
interactions between the particles forming the jets and the
detectors~\cite{Gross:2022hyw}. Another important task is to measure
the collimated hadronic decay of high-momentum heavy particles such as
top quarks, $W$, $Z$ or Higgs bosons. These are reconstructed as jets
with large angular opening (\largeR jets) and the measurement of
their mass as well as their energy is therefore necessary.
 
To measure the jet energy and mass from raw detector signals, the ATLAS and CMS
collaborations use calibration procedures
comprising multiple steps~\cite{JETM-2018-05,CMS-DP-2021-033}. In these procedures, the main energy and mass corrections are evaluated numerically from simulated collision events. The corrections are evaluated
and applied in two distinct successive steps, in bins of a limited (usually two or three) number of
characteristic variables. However, calibrating both highly correlated
energy and mass while exploiting all shower variables and their
correlations in a single correction step is possible through
approaches based on deep neural networks (DNN).
This paper presents a full implementation developed by the ATLAS
Collaboration using 13~\TeV simulated collision events, overcoming several difficulties that are specific to the
calibration problem. This new calibration method aims at replacing the current simulation-based corrections of the energy and mass of \largeR jets. After a brief introduction to the ATLAS detector
(Section~\ref{sec:ATLAS}), the paper explains why jet energy and mass calibrations are necessary, reviews the current procedure in ATLAS and
motivates the new DNN approach (Section~\ref{sec:std_calib}). The samples used and the selections applied to them are described in Section~\ref{sec:MC-dataprep}. Section~\ref{sec:method} details the
specific methodology needed to construct and properly train the
network, while Section~\ref{sec:validation} discusses the validation and performance of the resulting
DNN calibration, comparing it with the current numerical
method. Finally Section~\ref{sec:prospects} identifies and discusses
limitations in this approach as well as some possible solutions.


\section{The ATLAS detector}
\label{sec:ATLAS}
 
The ATLAS detector~\cite{PERF-2007-01} covers nearly the entire solid angle around the collision point.\footnote{\AtlasCoordFootnote}
It consists of an inner tracking detector surrounded by a thin superconducting solenoid, electromagnetic and hadron calorimeters,
and a muon spectrometer incorporating three large superconducting air-core toroidal magnets.
 
The inner-detector system (ID) is immersed in a \qty{2}{\tesla} axial magnetic field
and provides charged-particle tracking in the range of \(|\eta| < 2.5\).
The high-granularity silicon pixel detector covers the vertex region and typically provides four measurements per track, the first hit normally being in the insertable B-layer (IBL)~\cite{ATLAS-TDR-19,PIX-2018-001} installed before Run~2 of the LHC.
It is followed by the silicon microstrip tracker (SCT), which usually provides eight measurements per track.
These silicon detectors are complemented by the transition radiation tracker (TRT),
which enables radially extended track reconstruction up to \(|\eta| = 2.0\).
The TRT also provides electron identification information
based on the fraction of hits (typically 30 in total) above a higher energy-deposit threshold corresponding to transition radiation.
 
The calorimeter system covers the pseudorapidity range of \(|\eta| < 4.9\).
In the region \(|\eta|< 3.2\), electromagnetic calorimetry is provided by barrel and
endcap high-granularity lead/liquid-argon (LAr) calorimeters,
with an additional thin LAr presampler covering \(|\eta| < 1.8\)
to correct for energy loss in material upstream of the calorimeters.
Hadron calorimetry is provided by the steel/scintillator-tile calorimeter,
segmented into three barrel structures with \(|\eta| < 1.7\), and two copper/LAr hadron endcap calorimeters.
The solid angle coverage is completed with forward copper/LAr and tungsten/LAr calorimeter modules
optimised for electromagnetic and hadronic energy measurements respectively.
 
The angular and longitudinal segmentation of the calorimeters provides valuable 3-dimensional information about the location of the energy deposits. This allows the calculation of quantities useful for the characterization of the energy shower resulting from various type of particles or groups of particles such as hadronic jets.
 
An extensive software suite~\cite{ATL-SOFT-PUB-2021-001} is used in data simulation, in the reconstruction
and analysis of real and simulated data, in detector operations, and in the trigger and data acquisition
systems of the experiment.


\section{Jets and their calibration with the ATLAS detector}
\label{sec:std_calib}
 
After the parton emission, the complete hadronisation process
results in a set of particles from which the \emph{particle jets} are
formed. These are also named \emph{true} jets in the context of
simulated events discussed in this paper.  In practice the group of
particles forming a true jet, its constituents, is obtained by
applying a physically motivated procedure called a jet clustering
algorithm on the set of kinematic 4-vectors of all the stable interacting
particles of the collision (details are given in Section~\ref{subsec:montecarlo}).  The sum
of the 4-vectors of a jet's constituents defines its momentum and thus
the \emph{true} jet energy and mass.  These are the reference quantities for the jet
calibration that is applied to experimental jets reconstructed with the ATLAS detector.
Jet reconstruction is based on two types of primary
signals: charged particle tracks, which are reconstructed from
measurements of hits in the inner detector~\cite{IDET-2010-01,PERF-2015-08} and topological clusters,
which are reconstructed from energy deposits in the
calorimeters~\cite{PERF-2014-07}. Tracks and clusters can then be combined to form higher-level objects, each
characterised by a kinetic 4-vector~\cite{PERF-2015-09,ATL-PHYS-PUB-2017-015}. Together,
these objects approximate the flow of hadronic particles and are used as the constituents of the
reconstructed jets. The final jets are obtained by regrouping these
constituents using the same jet algorithm as used for the true jets.
 
Due to various detector inefficiencies and limitations, the kinematics of the
experimentally reconstructed jets differ significantly from the
corresponding true jets. Precise and complex corrections are
thus applied to the jet energy and mass before these quantities are used in
physics analyses~\cite{JETM-2018-05}.
These corrections are separated into two steps. Firstly, a series of
corrections is obtained from simulation that allows alignment between the reconstructed-level energies and masses
and the generated-level ones; they ensure that the jet energy scale (JES) or
the jet mass scale (JMS) of reconstructed jets is correct (a more precise
definition is given below). They also compensate for the effect of multiple proton--proton
interactions happening in the same bunch crossing and residual effects in the detector from
collisions in nearby crossings (these effects are collectively referred to
as \textit{\pileup}). Secondly, other corrections obtained from real
data, using known physical processes or well-reconstructed objects
such as photons, allow corrections for residual differences (of the
order of 2--3\%) to be made between real and simulated reconstructed jets. The calibration method presented in this paper is an alternative simulation-based correction.

Simulation-based corrections are derived from studies of the
\emph{individual response} of the quantity of interest $E$, the jet
energy, (but all the following equations stand identically for the
mass $m$), defined by $\rE=\frac{\Ereco}{\Etruebis}$ where
\Etruebis~ is calculated from the jets clustered
from generated particles  and \Ereco, from the jets reconstructed
after detector simulation ($\true$ and $\reco$ jets are more precisely
defined in Section~\ref{subsec:dataprep}).
Because of the stochastic nature of both the quantum chromodynamics (QCD) shower and its interactions with the
detector, true jets that have the same parameters (such as energy, angular
position, mass, \dots) can generate distributions of possible
\rE. Individual response distributions  depend on these parameters and are
usually unimodal:  their most probable value, their mode $\RE=\mode(\{\rE\})$, is
called the \emph{average response}, JES (or JMS in the equivalent case
for jet mass), or simply \emph{response} when
the context is non-ambiguous. The mode, as opposed to the mean or the
median, is chosen as the definition of such a central value to avoid
ambiguities or biases that can arise when the distributions are
asymmetric. The width of the distribution around the
mode is used to define the resolution in $E$. In this paper, the
chosen definition is the inter quantile range (IQR) at 68\%\footnote{IQR instead of the
standard deviation is chosen here as the shape of the response distribution does
not always follow a Gaussian distribution, particularly for the jet mass.} divided by the
mode $\sigma_E = \frac{\text{IQR}_{68\%}(\{\rE\})}{\RE} $. Because of
experimental limitations, the average response may differ from unity and
the objective of the calibration is to find the relevant corrections
such that $\RE=1$ for all populations of true jets (which formalizes the idea that
$\true$ and $\reco$ scales are \emph{on average} identical), with $\sigma_E$ as small as possible.

The individual response distributions can be
parameterised by a set of parameters describing the true jet (here taken to be
the energy, pseudorapidity and mass) but also
by inter-dependent parameters of the reconstructed jets such as the
energy, pseudorapidity, mass, fractions of energy deposited in the various
layers of the calorimeter, fraction of the momentum of the associated
tracks, and others.
Formally, it is then possible to model the distribution as a
probability distribution function $P(\rE \; | \; \vec{x})$ where
$\vec{x}$ is the list of input parameters that can be separated into the
various true and reconstructed parameters $\vec{x}=(\xtrue,\xreco)$.
The average response for true jets with parameters $\xtrue$ is then
given by
\begin{equation}
\label{eq:Rtrue}
\RE(\xtrue) = \mode(P(\rE \; | \; \xtrue)).
\end{equation}
The aim of the calibration is then to find a  function of the reconstructed
parameters $C(\xreco)$ defining the calibrated quantity
$E_\calib=C(\xreco)\Ereco$ such that the most probable value becomes
 
\[R_{E_\calib}(\xtrue) = \mode( P(\rE \times C(\xreco) \; | \; \xtrue)) = 1 \]
 
where $P(\rE \times C(\xreco))$ is the
probability distribution function obtained from $P(\rE)$ by changing $\rE \to r_{E_\calib} = \rE \times C(\xreco)$ and the mode
is taken over all $\xreco$.
 
In general this mathematical problem is difficult. It has an exact
solution in simple cases where the response depends on only one
parameter, the quantity $E$ itself, and when $\RE(\Etruebis)\Etruebis$ is an affine
function~\cite{Cukierman:2016dkb}. In this case the solution is to choose the
function $C$ such as
\begin{equation}
\label{eq:nisolution}
\forall x \in \mathbb{R}, \quad C(x\RE(x)) = \frac{1}{\RE(x)}.
\end{equation}
This solution is called the \emph{numerical inversion} (NI) because $\RE$ (and thus
the pair $(x\RE(x),\frac{1}{\RE(x)})$) can be obtained \emph{numerically} from
the simulated data and because $\Etruebis \RE(\Etruebis)$ represents an
\emph{inversion} of the true quantity.\footnote{For a given $E_\true$,
the most probable $\Ereco$ is by definition $\Etruebis \RE(\Etruebis)$.}
This method (referred to here as \emph{standard calibration}) is
employed by ATLAS to obtain sequentially a jet energy scale
calibration and then a mass  scale calibration.
In practice, in the standard calibration, $\RE$ is evaluated as a function of $(E_\true,\eta_\true)$
(and $m_\true$ for mass calibration) by collecting individual
responses in bins of these quantities. This function $\RE$ is then converted into a correction factor function depending only on reconstructed quantities ($E_\reco,\eta_\reco$ and $m_\reco$) according to the NI principles described above and further detailed in \cite{JETM-2018-05,PERF-2011-03}. This procedure
gives satisfactory performance. The variables $E$, $\eta$ and $m$ are the
principal dependencies of the JES and JMS and the use of true quantities
to form the response evaluation bins allows potential biases
due to the energy or mass spectrum of the generated sample to be avoided when
evaluating the average response.
 
As mentioned above, there is much more information available to
describe the reconstructed hadronic showers than just the energy, pseudorapidity
and mass. Using these additional variables can potentially help obtain a more
precise JES and JMS and improve their resolution, including
subsamples of jets such as quark-initiated or gluon-initiated jets. However, these
variables can be correlated. Optimally exploiting all the
correlations in the same way as performed in the standard calibration
would require highly multidimensional bins, which is practically impossible.
On the other hand the task of numerically modelling a
multidimensional function can be successfully performed with
modern DNN techniques. Furthermore, a key problem of the calibration evaluation, estimating the mode
of distributions, is solvable with these techniques; a solution is discussed in detail in Section~\ref{subsubsec:loss}.
A proof of concept of using DNN for JES and JMS calibration, including an implementation of the
NI procedure, has already been presented
in~\cite{ATL-PHYS-PUB-2018-013} and DNN techniques are also employed to perform a particular
step of the energy (but not the mass) calibration for small-\emph{R} jets~\cite{JETM-2022-01}. Building on this, a
complete and operational solution and its performance is presented
here. This new procedure extends several aspects of the DNN calibration
but does not implement NI to simplify the method by
avoiding the need to train two DNNs sequentially, one to model the
function $R$ as a function of $x_\true$, the second the function $C$
as in Eq.~(\ref{eq:nisolution}). Instead, it evaluates directly the
uncalibrated responses in terms
of reconstructed quantities, which means that the network evaluates
\begin{equation}
\label{eq:Rreco}
\RE(\xreco) = \mode(P(\rE \; | \; \xreco))
\end{equation}
instead of $\mode(P(\rE \; | \; \xtrue))$ as in
Eq.~(\ref{eq:Rtrue}). The calibration function is then chosen to be
simply $1/\RE(\xreco)$. The price to pay for this simpler procedure is that the
response $\mode(P(\rE \; | \; \xreco))$ a priori depends on the
distribution of the simulated input samples and this dependency can bias the
resulting calibration even in the simple cases where the standard calibration,
which uses $\mode(P(\rE \; | \; \xtrue))$, would not be biased.
However, because $\xreco$ includes many
variables relevant to the description of the hadronic shower, it can
be expected that the response distributions at each point of the phase space are
narrow enough so that the overall bias on the mode is negligible. Although this hypothesis
is not testable in practice because of the high dimensionality,
verifications are performed to ensure the input samples do not
bias the performance (they are presented in Section~\ref{subsubsec:spectrumdep}).


\section{Samples and selections}
\label{sec:MC-dataprep}
 
\subsection{Event simulation}
\label{subsec:montecarlo}
 
The training and validation of the DNN is based on Monte Carlo
simulations of proton--proton (\(pp\)) collisions at a centre-of-mass energy of
$\sqrt{s}=13$~\TeV. All events were passed through the complete ATLAS
detector simulation based on \GEANT \cite{AGOSTINELLI2003250,
SOFT-2010-01}. Various physics processes were simulated as described
in Table~\ref{tab:dataset} to train and test the DNN on different jet
topologies.
 
\begin{table}[!htp]
\centering
\caption{Simulated samples used for training and validation}
\begin{tabular}{lllc}
Use & Process type & Generator & Number of jets passing the selections (in millions) \\
\hline
Training, validation & QCD dijet & \PYTHIA[8.230] & $\sim$ 270 \\
Validation           & $W/Z$  & \PYTHIA[8.230] & $\sim$ 20 \\
Validation           & Top quark& \PYTHIA[8.230] & $\sim$ 15 \\
Validation           & Higgs Boson & \PYTHIA[8.230] & $\sim$ 15 \\
Validation           & QCD dijet & \SHERPA[2.2.5] & $\sim$ 30
 
\end{tabular}
\label{tab:dataset}
\end{table}
Training is performed on the QCD dijet sample as this type of process is,
by orders of magnitude, more probable at the LHC and enters as a
background to many analyses. In a validation step, it is verified that the
training generalises and produces a good calibration for other important
physics processes such as those listed in Table~\ref{tab:dataset}. All
samples (except the \SHERPA one) were produced using \PYTHIA[8.230]~\cite{Sjostrand:2014zea} with the \NNPDF[2.3lo]~\cite{Ball:2012cx}
parton distribution function (PDF) set and the A14 set of tuned parameters~\cite{ATL-PHYS-PUB-2014-021}. The training sample containing jets
originating from light quarks or gluons was obtained by generating
dijets in slices of the jet transverse momentum, \pT, to sufficiently
populate the kinematic region of interest (from around 200 \GeV up to 6000
\GeV).
For the validation samples, physics processes modelling phenomena
beyond the Standard Model are used to populate the high-\pT
region.
\begin{itemize}
\item The $W/Z$-boson sample was obtained by generating the $W' \to WZ \to q\bar{q}q\bar{q} $ process with $m_{W'} =2$~\TeV.
\item The top-quark sample was obtained by generating the $Z' \to t \bar{t} \to W^+b \; W^-\bar{b} \to bq\bar{q} \; \bar{b}q\bar{q}$ process with $m_{Z'} =4$~\TeV.
\item The sample of Higgs bosons, which displays a broad Higgs boson \pT spectrum, was obtained by generating the production of Randall--Sundrum gravitons $G^*$ in a benchmark model with a warped extra dimension~\cite{Randall_1999} over a range of graviton masses from 300 to 6000~\GeV, where the graviton decays according to $G^* \to HH \to b\bar{b}b\bar{b}$.
\end{itemize}
 
The \SHERPA dijet sample was generated using the \SHERPA[2.2.5]~\cite{Bothmann:2019yzt} generator.
The matrix element calculation was included for the \(2\rightarrow2\) process at leading order, and the default \SHERPA parton shower~\cite{Schumann:2007mg} based on Catani--Seymour dipole factorisation was used for the showering with \pT ordering, using the \CT[14nnlo] PDF set~\cite{Dulat:2015mca}. These samples used the dedicated \SHERPA AHADIC model for hadronisation~\cite{Winter:2003tt}, based on cluster fragmentation ideas.
 
Finally, the effect of multiple interactions in the same and neighbouring bunch
crossings (\textit{\pileup}) was modelled by overlaying the simulated
hard-scattering event (event of interest) with inelastic \(pp\) events generated with \PYTHIA[8.186]~\cite{Sjostrand:2007gs} using the \NNPDF[2.3lo] set of PDFs~\cite{Ball:2012cx} and the
A3 set of tuned parameters~\cite{ATL-PHYS-PUB-2016-017}. The number of
overlaid events is representative of the amount of pile-up present in
the 2017--2018 ATLAS $pp$ data sample.

\subsection{Jet reconstruction and selection}
\label{subsec:dataprep}
 
In each event, the flow of hadronic particles is represented by a set
of 4-vectors built by combining momentum and energy measurements from
the inner tracker and
calorimeters.
In this paper, the combination technique producing
\emph{unified flow objects} (UFO)~\cite{JETM-2018-06} is used. These input constituents are
clustered into jets with the \antikt jet algorithm with angular radius
$R=1.0$~\cite{Cacciari:2008gp} using the \textsc{FastJet}
software package~\cite{Cacciari_2012}. The jets then undergo a
\emph{grooming} procedure aimed at suppressing soft radiation from
pile-up and the underlying event; the \emph{soft-drop}
procedure~\cite{Dasgupta:2013ihk} is employed with
parameters $\beta = 1$ and $z_\mathrm{cut}=0.1$.
 
Stable particles, with a lifetime $c\tau>10$ mm, produced by the generators (with the exception of neutrinos and muons) are also clustered into jets using the same
algorithm to obtain \emph{true} jets.
 
To obtain training samples that are not biased by experimental effects
such as pile-up, noise or misidentifications, some requirements are
imposed on the jets. Reconstructed jets are selected if they are
matched to a true jet and isolated as described in
Ref.~\cite{JETM-2018-05}.  A reconstructed and a true jet are matched
if they are within an angular distance $\Delta R=0.3$. Two isolation
requirements are then applied; reconstructed jets are rejected if a
true jet other than the matched one is found within $\Delta R = 2.5$
or if another reconstructed jet is found within $\Delta R=1.5$.
Finally, both the reconstructed and true jets are required to have
$\pt>\qty{50}{\GeV}$ and, in an event, only the two highest-\pT jets
satisfying these criteria are considered.  The selection results in about
270 million jets being available for the DNN training sample.


\section{Methodology of the deep neural network calibration}
\label{sec:method}
 
The simultaneous calibration of the jet energy and mass is performed using
the calculations of a single DNN taking as
inputs reconstructed jet-level and event-level quantities. The DNN is set up to output two numbers
corresponding to the energy and mass responses. However,  as mentioned in
the previous section, it is not possible to directly
predict the exact individual $E$ (or $m$) response for a given reconstructed jet as each
true jet corresponds to a distribution of possible responses.
The DNN predictions are thus interpreted as the most probable values
of the possible responses for the input jets (that is, the modes of
their distribution). Following Eq.~(\ref{eq:Rreco}), the DNN
is trained so that
 
\begin{equation*}
R^{\dnn}_E(\xreco)
\simeq \RE(\xreco) \equiv \mode(P(\rE \; | \; \xreco)).
\end{equation*}
 
The calibrated energy (or mass) is then defined as
 
\[ E_\calib =  \frac{\Ereco}{R^{\dnn}_E(\xreco)}.\]
 
Thus, the DNN output predictions are
 
\[ \ypred = (R^\dnn_E, R^\dnn_m). \]
 
During the training, the target quantities corresponding to the inputs
$\xreco$ are the individual responses discussed in Section~\ref{sec:std_calib}:
 
\[ \vec{y}_{\text{target}}= \ytrue = (r_E, r_m) = (\frac{E_\reco}{E_\true}, \frac{m_\reco}{m_\true}).\]
 
These are not directly the desired \emph{average} responses $R_E$ and $R_m$ and the
necessity to predict distribution modes (and not means nor medians) requires a careful choice of the
loss function.
Beside these mathematical concerns, experimental difficulties related
to the detector geometry -- and especially the geometry of the
calorimeters -- strongly interfere in the calibration predictions for certain angular regions. This is
accounted for by a dedicated processing of the jet angular
position.

Finally, the mass calibration alone has specific
difficulties. Firstly, the mass response distributions are generally
wider and asymmetric, making it harder for the DNN to predict their
mode. Secondly, \largeR jets are mainly used to reconstruct heavy
particles whose decays are collimated (top quarks, $W$, $Z$, or Higgs
bosons) and for which the mass measurement is crucial.
So the DNN performance needs to be good for these jets, even if their topology differs from standard QCD jets.
This has motivated the use of a specific architecture (including a
\emph{residual connection}-like structure)
and training procedure (with a dedicated mass-only step).

\subsection{Loss function}
\label{subsubsec:loss}
 
The loss function $\mathcal{L} (\ytrue,\ypred)$
is a crucial part of the DNN training as it defines what mathematical
quantity the DNN will learn to predict. This is particularly true in the case
presented here where the targets ($\ytrue$) are realizations of a stochastic
distribution. In this case, it is proven that minimising
the usual mean square error loss
($\mathcal{L}=(\ytrue-\ypred)^2$)  would have $\ypred$ converge
to the mean of the distribution, while minimising the mean absolute
error ($\mathcal{L}=|\ytrue-\ypred|$) would imply a convergence to
the median~\cite{Cheong:2019upg}. Converging to the distribution mode, as is
necessary for the jet calibration, would require taking a Dirac delta
distribution as the loss $\mathcal{L}=\delta(\ytrue-\ypred)$.
This is not possible in practice and an approximate function is therefore
necessary. One possibility is the Gaussian kernel as suggested in Ref.~\cite{Cheong:2019upg},
another solution is inspired by mixture-density-network (MDN) techniques
and is described below. Numerical tests show the MDN losses are less biased
than the Gaussian kernel losses. They are also found to converge faster
and allow more flexibility in the training procedure.
 
First, as the DNN produces two outputs $\ypred = (R^\dnn_E,
R^\dnn_m)$, two independent loss functions are considered, the
sum of them being the final minimised quantity:
 
\[ \mathcal{L}_{tot} = \mathcal{L}_E + \mathcal{L}_m. \]
 
The MDN loss is constructed from the assumption that the response
distribution of the DNN targets is a Gaussian distribution with central value $\mu$
and width $\sigma$.
\[P(\rE, (\mu,\sigma)) \propto \frac{e^{-(\rE-\mu)^2/2\sigma^2}}{\sigma}.\]
 
With the parameter of interest being
$\mu$ (the mode of the distribution), an estimator $\hat{\mu}$ (and
$\hat{\sigma}$) can be obtained from the maximum of the likelihood (LH)
over a given sample of $\rE$
\begin{equation}
\label{eq:lhP}
\mathrm{LH}(\mu, \sigma) = \prod_{\rE} P(\rE,  (\mu, \sigma) ).
\end{equation}
 
As a consequence, a DNN outputting two values $\ypred=(\mu_\pred,
\sigma_\pred)$ and trained with the loss:
 
\begin{equation}
\label{eq:mdnloss}
\mathcal{L}_{\text{MDN}} = -\log(P(\ytrue, \ypred )) = \log(\sigma_\pred)+\frac{1}{2} \frac{(\rE-\mu_\pred)^2}{\sigma_\pred^2}
\end{equation}
 
would minimise the logarithm of Eq.~(\ref{eq:lhP}) and thus predict the estimators,
that is $(\mu_\pred, \sigma_\pred)=(\hat{\mu}, \hat{\sigma})$.
Using this MDN loss, the DNN  has to predict the width of this distribution ($\sigma_\pred$) in
addition to predicting the mode of the response distribution. For an energy and mass calibration DNN:
 
\[ \ypred =  \begin{pmatrix} \mu^E_\pred \\ \sigma^E_\pred \\  \mu^m_\pred \\ \sigma^m_\pred \end{pmatrix},  \]
 
where only the predicted modes are eventually used in the final calibration procedures.

In practice, the energy or mass response distributions are not perfect
Gaussian distributions especially for jets with low \pT and high mass or with
high $|\eta|$. To take this into account, the assumed underlying
distribution can be replaced by an asymmetric Gaussian distribution or a truncated
Gaussian distribution.
\begin{equation}
P_{\textrm{asym}}(x) \sim \begin{cases} e^{(x-\mu)^2 / 2\sigma^2_1} &
\text{if }  x<\mu \\
e^{(x-\mu)^2 / 2\sigma^2_2} &    \text{if } x\geq\mu
\end{cases} \quad
P_{\textrm{trunc}}(x) \sim \begin{cases} e^{(x-\mu)^2 / 2\sigma^2} &
\text{if }  |x-\mu|<N \sigma \\
0 & \text{otherwise}
\end{cases}
\end{equation}
Although not physically motivated these losses help with the
numerical convergence of the network. In particular, the asymmetric MDN
loss (MDNA) helps to avoid biases from asymmetric shapes in the response
distributions, but it requires the DNN to output three values per prediction ($\mu,
\sigma_1, \sigma_2$). The use of a truncated distribution can apply to
both the MDN and MDNA cases. In practice, truncation consists of excluding
from the loss evaluation jets for which $|x-\mu| >N \sigma$ and
adapting Eq.~(\ref{eq:mdnloss}) with the relevant distribution
normalisation factor. Asymmetric loss helps to converge faster in
regions of phase space where the distributions are asymmetric and the
truncation is used at the end of the training to improve the accuracy
of the calibration.  Details of the choice and tuning of the
parameters of the loss function are further discussed in Section~\ref{subsec:training}.

\subsection{Input features}
\label{subsubsec:features}
 
To obtain precise predictions, it is necessary to give the DNN
an as complete description of the jet and the event as possible. A
list of 21 jet or event variables is chosen as the input to the
DNN\@. These features are chosen for their relevance in calibration or
jet identification tasks as demonstrated in previous
studies~\cite{JETM-2018-05,JETM-2018-03}. They can be divided in four categories:
the jet kinematics, which are the main dependencies of the energy and mass
responses;\footnote{Although the rapidity provides a better estimate of the actual
angular position of massive objects, the pseudorapidity is used in
this paper as used in the current ATLAS calibration; the performance
is not expected to be affected by this choice} the jet substructure
variables, which describe the particle-shower topology
used to classify jets; the detector
variables, which quantify the relative charged-particle contribution
and the development of the calorimeter showers; and
finally, the event variables that describe the jet environment. Table
\ref{tab:Features} presents these input variables and their
definitions.
To have homogeneous inputs for the DNN, each input is
normalised so that its distribution falls in the range of $[-1,1]$.
 
\begin{table}[!htp]\centering
\caption{The input features of the DNN.}
\scriptsize
\begin{tabular}{lrlr}\toprule
&Name &Definition \\\cmidrule{2-3}
\multirow{3}{*}{Jet level} & $E$ & Energy of the jet in \GeV, $\log E$ is taken to reduce the spread of its distribution  \\
& $m$ & Mass of the jet in \GeV, $\log m$ is taken to reduce the spread of its distribution \\
&$\eta$ & Jet pseudorapidity \\\cmidrule{1-2}
\multirow{5}{*}{Substructure level}        &groomMRatio & Mass ratio of groomed to ungroomed jets \\
& Width &  $\sum_i p_{\mathrm{T}i}\Delta R(i,\text{jet}) /(\sum_i
p_{\mathrm{T}i})$ where $\Delta R$ is the angular distance (sum over the jet constituents) \\
&Split12,Split23 & Splitting scales at the 1st and 2nd exclusive $k_T$ declusterings~\cite{Thaler:2008ju}  \\
&C2, D2 & Energy correlation ratios~\cite{Larkoski:2013eya,Larkoski:2014gra} \\
&$\tau_{21}$, $\tau_{32}$ & N-Subjettiness ratios using WTA axis~\cite{Thaler:2010tr,Larkoski:2014shape} \\
&Qw & Smallest invariant mass among the proto-jets pairs of the last 3 steps of a $k_T$ reclustering sequence\\
\cmidrule{1-2}
\multirow{7}{*}{Detector level} &EMFrac & Energy fraction deposited in the electromagnetic calorimeter \\
&EM3Frac & Energy fraction deposited in the third layer of the electromagnetic calorimeter \\
&Tile0Frac & Energy fraction deposited in the 1st layer of the
hadronic calorimeter \\
&EffNConsts & $(\sum_i E_i)^2  / (\sum_i E_i^2)$ (sum over the jet constituents)\\
&NeutralFrac & Energy fraction from neutral constituents \\
&ChargedPTFrac & \pT~fraction from charged constituents \\
&ChargedMFrac & Mass fraction from charged constituents\\\cmidrule{1-2}
\multirow{2}{*}{Event level} &$\mu$ &  Mean number of interactions per bunch crossing \\
&NPV & Number of primary vertices per event \\\midrule
\bottomrule
\end{tabular}
\label{tab:Features}
\end{table}

In addition, the DNN has difficulties learning the sharp variations in
the energy and mass responses along $\eta$ that are due to the complex
ATLAS detector geometry and in particular
to the poorly instrumented regions between the central and forward
calorimeters between $|\eta|=0.8$ and $|\eta|=1.5$.

\begin{figure}[htbp]
\begin{center}
\begin{subfigure}[c]{0.25\textwidth}
\centering
\includegraphics[width=\textwidth,keepaspectratio]{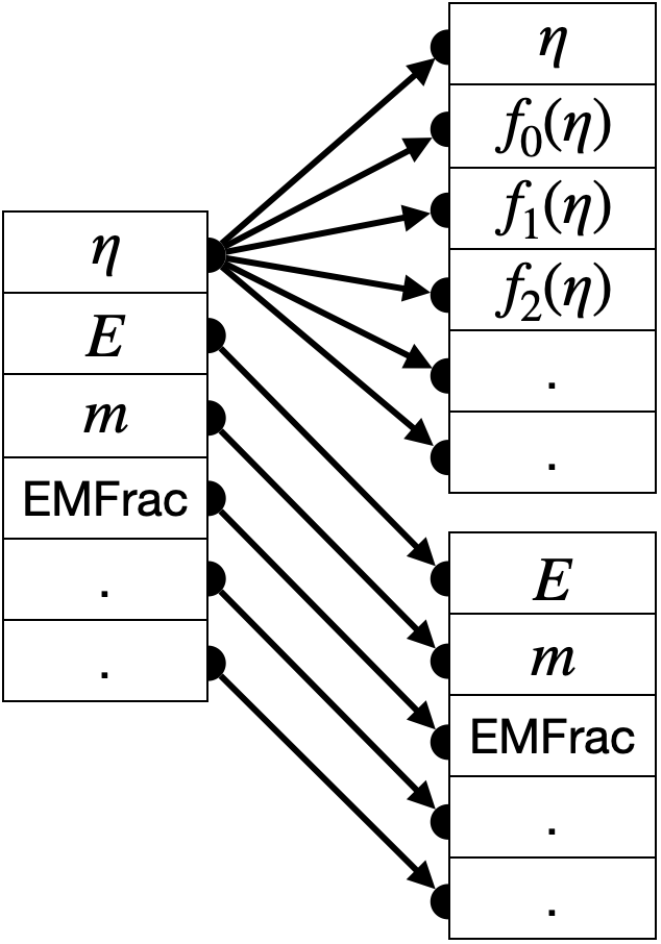}
\end{subfigure} \hfill
\begin{subfigure}[c]{0.6\textwidth}
\centering
\includegraphics[width=\textwidth,keepaspectratio]{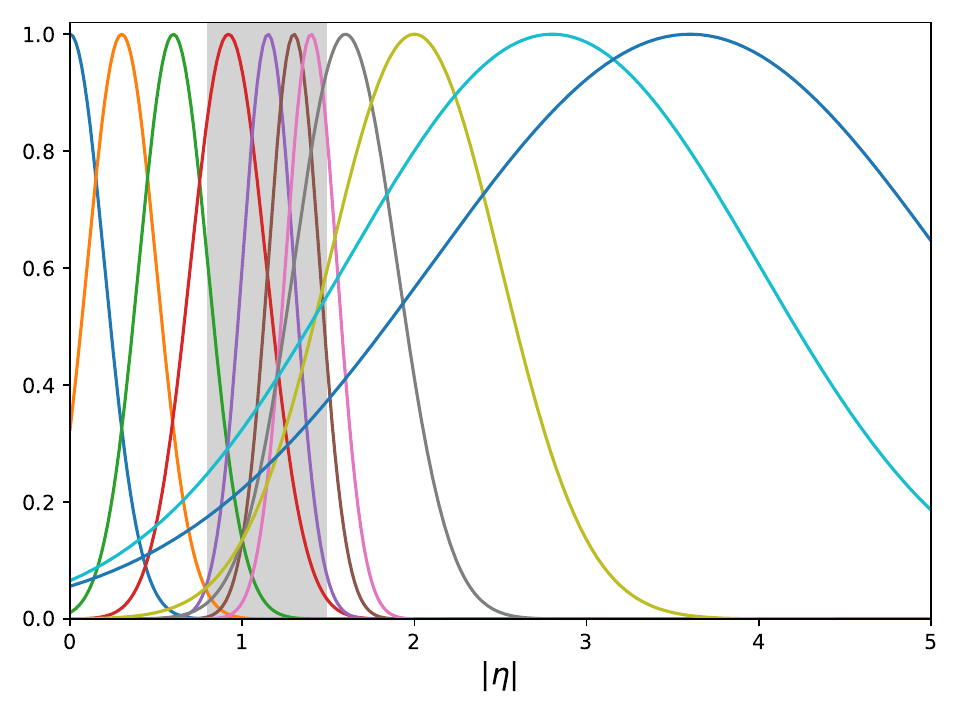}
\end{subfigure}
 
\end{center}
\caption{The principle of the $\eta$ annotation (left) and the various $f_i(\eta)$ functions (right). The vertical shaded band on the right plot marks the calorimeter transition region.}
\label{fig:EtaAnnot}
\end{figure}

In an effort to help the modelling, an additional input
preparation step called \emph{$\eta$ annotation} is
implemented as the first operation of the DNN\@. This step computes
extra features based on $\eta$ that are passed along with other
input features to the rest of the DNN\@. These extra features are calculated
as $f_i(\eta)$ where $f_i$ are Gaussian functions with different
central values and widths that are chosen to cover key angular regions
of the detector (such as the central part or edges of the electromagnetic barrel calorimeter and of other calorimeters). Figure~\ref{fig:EtaAnnot} shows how the 11 different
Gaussian functions are positioned, providing a continuous
encoding of the proximity of the jet to these regions. These processed values are included alongside the unprocessed $\eta$ for a total of 12 representations.
 
\subsection{Architecture}
 
\begin{figure}[htbp]
\begin{center}
\includegraphics[width=14cm,keepaspectratio]{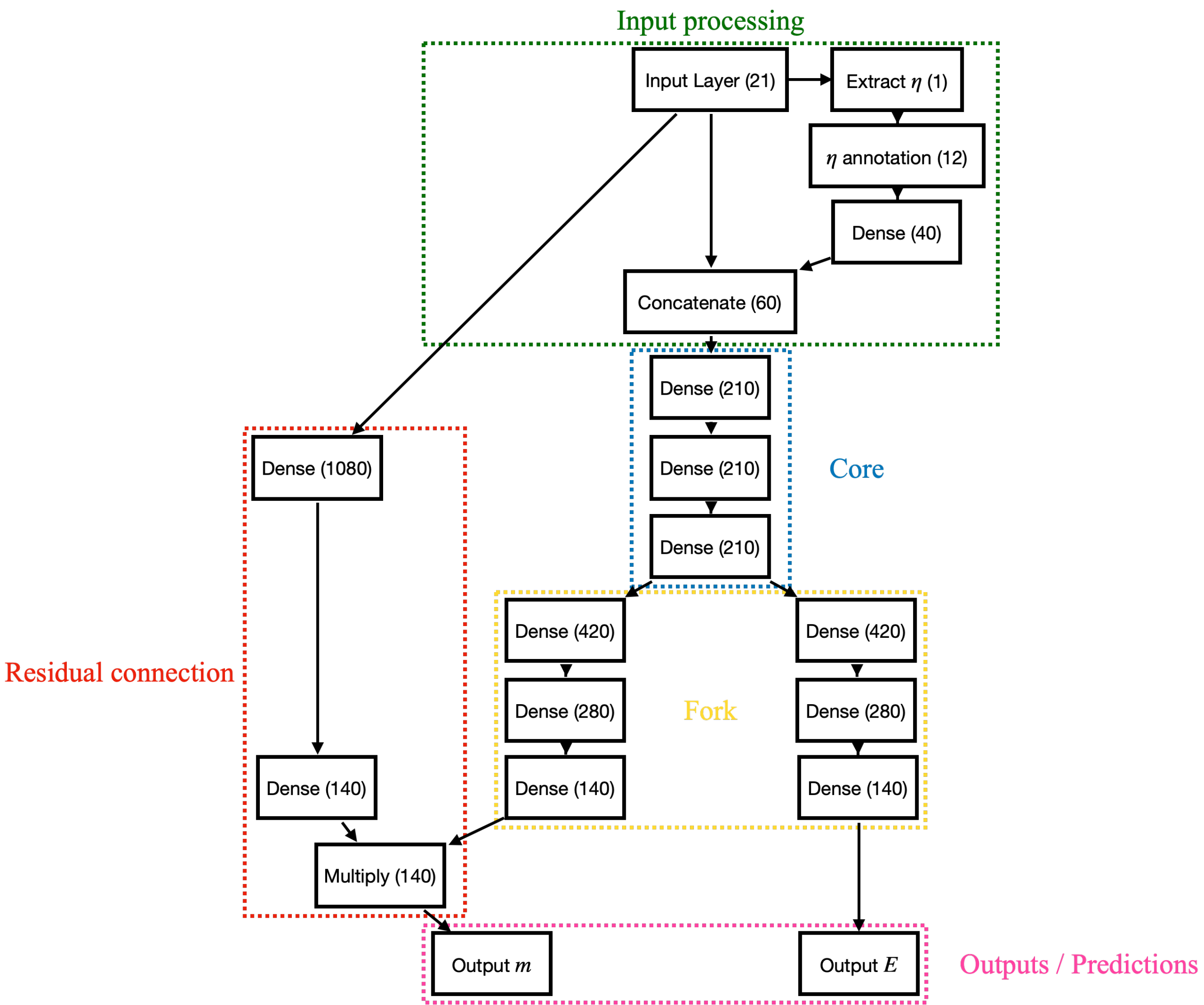}
\end{center}
\caption{The architecture of the DNN.}
\label{fig:architecture}
\end{figure}

The DNN architecture is mainly based on two specific parts, the $\eta$ annotation (described in Section~\ref{subsubsec:features}) and a fork between the energy and mass outputs. Both are designed to address the main difficulties of the jet calibration, the $\eta$ dependency of the responses and the important differences between the energy and mass response distributions. Building on this base, the architecture was iteratively improved and tested to obtain a structure giving good performance. Consequently, the DNN has a relatively complex structure as represented in Figure~\ref{fig:architecture}, where:
\begin{itemize}
\item \emph{Dense (N)} blocks represent the matrix, bias and activation
function resulting in a hidden layer with N neurons.
\item The \emph{Concatenate} block is a static layer (without trainable
parameters) that concatenates its two input layers.
\item The \emph{Multiply} block is a static layer that performs an
element-wise multiplication of its input layers. 
\item The final \emph{Output} layers give the predictions for the
energy and mass responses. They may consist of two or three
neurons depending on the loss function (as discussed in Section~\ref{subsubsec:loss}).
\end{itemize}
 
The DNN contains approximately 764~000 trainable
parameters and is structured in different parts as indicated in
Figure~\ref{fig:architecture}.:
\begin{itemize}
\item Input processing: contains the input layer with the 21 input features and the $\eta$ annotation described above.
\item Core: contains several hidden layers (210 neurons each) common to both the energy and mass predictions.
\item Early fork: premature splitting between the energy and mass
predictions to allow for a strong specialisation of the DNN in
each calibration, both forks containing the same hidden
layers. Having independent weights also allows to \emph{freeze} some
of them during the training procedure.
\item Multiplicative residual connection: loosely inspired from Ref.~\cite{kaiming2015} with
the substitution of the addition by a multiplication to increase the class of functions that can
be modelled by the network.
These layers link the input layer directly to
the mass output, with the intention of making the DNN learn which
inputs are the most important for the mass calibration and thus
helping to improve the mass predictions.\footnote{This is only added for the mass calibration prediction as it is harder to predict than the energy calibration due to its response distributions being broader and more asymmetric. The predicted energy calibration already shows excellent performance without an additional residual connection.}
\item Outputs: the DNN has two output predictions, one for the energy calibration and one for the mass calibration, $\ypred = (\mu_{\pred}^E, \mu_{\pred}^m) = (R^{DNN}_E, R^{DNN}_m)$, along with subsidiary predictions ($\sigma^E_\pred, \sigma^m_\pred $).
\end{itemize}
 
The activation function for each hidden layer is chosen to be
the MISH function~\cite{mish} and is given by:
 
\[f(x) = x \tanh(\ln(1+e^{x})).\]
 
The only exception is the activation function of the last hidden layer of the
residual connection that is the \emph{softmax} function: its outputs
are restricted to $[0,1]$ and are used as multiplicative weights.
 
The last activation function applied to the output layers is chosen to
be:
 
\[f(x) = \tanh(x) +1\]
 
to obtain predictions  between $0$ and $2$, which is
the expected range of the energy and mass central responses.

\subsection{Training procedure}
\label{subsec:training}
 
Establishing the training strategy cannot be left to a simple
hyperparameter scan. Firstly, because the energy  and mass responses
vary significantly across the phase space that is
not uniformly populated and, secondly, because converging
to a loss minimum does not guarantee that the physics performance goals are
satisfied for the full phase space. Furthermore, these performance goals
(described in Section~\ref{sec:validation}) are costly to evaluate and
cannot be tested during the training procedure. As a consequence the strategy
presented here is obtained with a series of attempts and
refinements gradually improving the performance.
This training strategy is described in
Table~\ref{tab:DNNtraining}: it totals 107 epochs and is divided into three parts.

In a first initialisation step, the DNN is trained with small batch sizes
and an MDNA loss with no or small truncation of the
loss. This aims at quickly converging to a rough estimate of the mode
of the response distributions thanks to small batch size (i.e.,
frequent weight updates) and a few epochs while adapting to possible
asymmetric distributions (e.g., at high $\eta$) with the MDNA loss.
 
The second step is designed to obtain very precise
predictions by taking advantage of large batch sizes and a progressive
narrowing of the loss truncation. This narrowing allows a gradual
focus on the centre of the response distributions, improving
the predictions. The larger batch sizes improve the calculation speed and
thus the training time.
 
The third step is a fine-tuning step with an exclusive mass
training. The weights involved in the energy prediction\footnote{These comprise the weights involved
in the input processing, the core and the energy-dedicated fork of the network architecture.} are frozen to
their values at the end of the second step and only those involved in
the mass prediction\footnote{These comprise the weights in the mass-dedicated fork and in the
residual connection layers of the network architecture.} are free to be tuned.
This step is necessary in order to have as good predictions for the mass as for the
energy for which the two first steps are enough. For this last step, a
MDN loss with a large truncation at $1.0\sigma$ is used in order to
focus more on the core of the mass response distribution.
A symmetric loss at this stage is observed to converge better,
possibly because it results in a simpler network, easier to optimize on
the central and more symmetric parts of the response distributions.
 
During the training two different optimizers are used, \emph{Rectified Adam} \cite{radam} in the initialisation step and \emph{diffGrad} \cite{diffgrad} in the common training and exclusive mass training. The learning rate is decreased
in the latter steps of the training following the improvements in the predictions.
 
\begin{table}[!htp]\centering
\caption{The specific training procedure of the DNN\@. MDN(A) stands for mixture density network (asymmetric). The loss truncation is indicated in number of standard deviations, $\sigma$. When $E$ and $m$ are not mentioned, it means that the energy and mass losses are identical.}
\scriptsize
\begin{tabular}{lcrrrlr}\toprule
Steps &N° &Number of epochs &Batch size &Learning rate &Loss \\\cmidrule{1-6}
\multirow{4}{*}{Initialisation} &1 &2 &15~000 & $10^{-3}$ &MDNA \\\cmidrule{2-6}
&2 &2 &25~000 & $10^{-3}$ &MDNA \\\cmidrule{2-6}
&3 &2 &35~000 & $10^{-3}$ &MDNA truncated ( $4.0\sigma$ ) \\\cmidrule{2-6}
&4 &2 &15~000 & $10^{-3}$ &MDNA truncated ( $3.5\sigma$ ) \\\cmidrule{1-6}
\multirow{6}{*}{Common training} &5 &6 &95000 & $10^{-3}$ &MDNA truncated ( $3.5\sigma$ ) \\\cmidrule{2-6}
&6 &6 &95~000 & $10^{-3}$ &MDNA truncated ( $3.5\sigma$ ) \\\cmidrule{2-6}
&7 &6 &125~000 & $10^{-3}$ &MDNA truncated ( $3.2\sigma$ ) \\\cmidrule{2-6}
&8 &6 &125~000 & $10^{-3}$ &MDNA truncated ( $3.2\sigma$ ) \\\cmidrule{2-6}
&9 &10 &155~000 & $5.10^{-4}$ &MDNA truncated ( $3.0\sigma$ ) \\\cmidrule{2-6}
&10 &15 &95~000 & $10^{-5}$ &MDNA truncated ( $E$: $3.0\sigma$, $m$: $2.0\sigma$ ) \\\cmidrule{1-6}
Exclusive mass training &11 &50 &95~000 & $10^{-5}$ &MDN truncated ( $1.0\sigma$ ) \\\midrule
\bottomrule
\end{tabular}
\label{tab:DNNtraining}
\end{table}


\section{Performance and validation of the calibration}
\label{sec:validation}
 
The evaluation of the DNN performance and the validation of the predicted energy and mass calibrations are
presented in this section. Unless mentioned otherwise, these are evaluated on the dijet sample,
hence for standard quark and gluon jets.

\subsection{Technical performance}
 
\begin{figure}[htbp]
\begin{center}
\includegraphics[width=12cm,keepaspectratio]{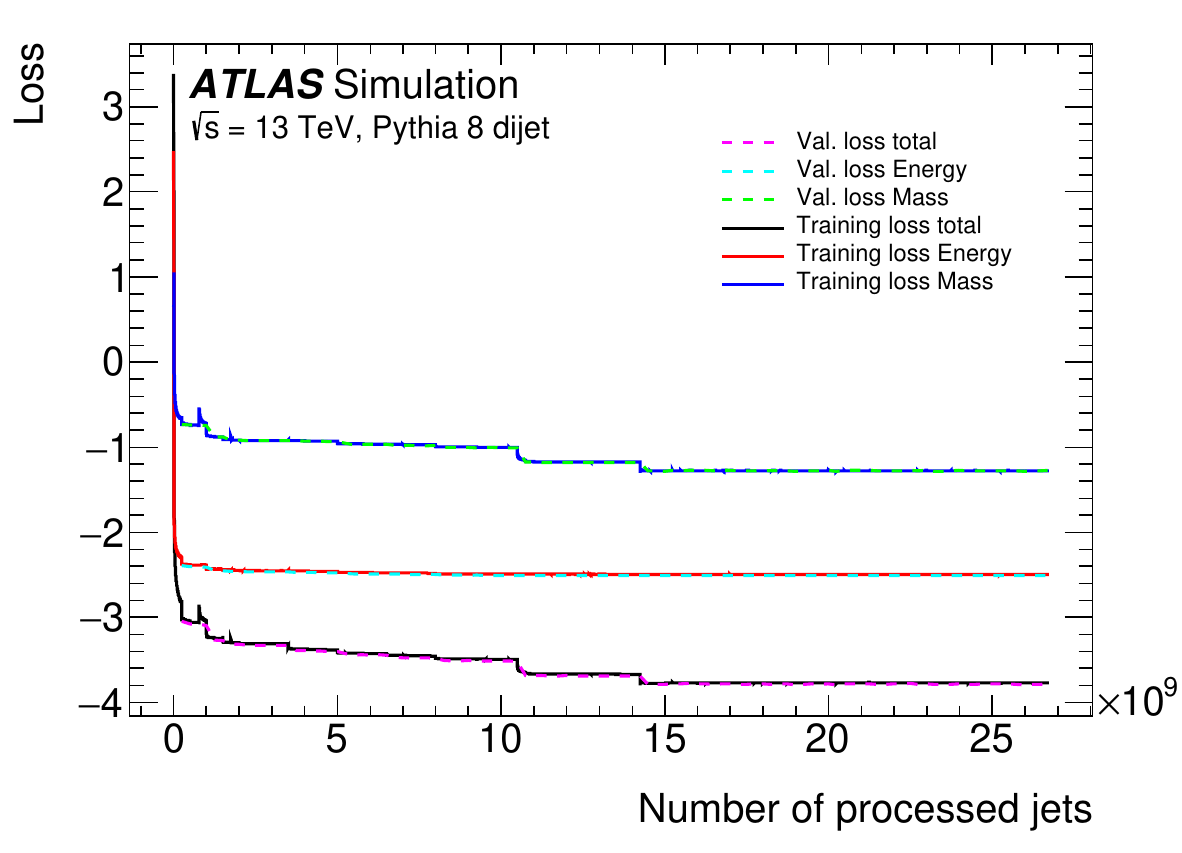}
\end{center}
\caption{The loss evolution for the training (plain lines) and
validation (dotted lines) data samples. The different spikes and
drops correspond to the transitions between training steps (see
Table~\ref{tab:DNNtraining}) and are due to the variations of the loss function.}
\label{fig:val_loss}
\end{figure}

For this study, the validation data sample is a subset of the dijet data sample that
is set aside at the beginning of the training (about ten million jets corresponding to $\sim4\%$
of the training data sample). Figure~\ref{fig:val_loss} shows the evolution of the loss with the
number of epochs (i.e., the number of jets processed). The training and validation losses (obtained for
the energy, the mass and their total) are shown as solid and dotted lines, respectively. The different
losses display very good agreement, showing that the DNN does not overfit the training data sample, as expected
given the size of the training data sample; overfitting  would take many more epochs.
It also shows that the loss does not evolve significantly after two thirds of the training. For a more common DNN
set-up, it could be wise to stop the training earlier, as it seems -- by that metric alone -- that its performance
is not improving anymore. However, it is observed that the predicted
calibrations continue to improve even if the loss does not, which makes it difficult to find an optimum training procedure.
 
The consistency of the training is also verified by training DNNs that
differ only by weight initialisation and ensuring that their losses and
calibration performance are identical.

\subsection{Calibration performance}
 
The performance of the energy and mass calibrations are quantified
using the response $R_{E_\calib}$ and the resolution $\sigma$ of the calibrated
distributions, as defined in Section~\ref{sec:std_calib}.
 
More precisely, from the predicted energy and mass response
$\ypred = (R_{E}^{\text{DNN}}, R_{m}^{\text{DNN}})$, the individual energy (or mass) calibrated response is defined as:
 
\[ r_{E_\calib} =  \frac{E_\calib}{E_\true} = \frac{1}{R_E^{\text{DNN}}} \frac{E_\reco}{E_\true} = \frac{r_{E}}{R_E^{\text{DNN}}}. \]
 
As discussed in Section~\ref{sec:std_calib}, $r_{E_\calib}$ values are
realisations of a distribution $P$ from which the calibrated response
can be obtained: $R_{E_\calib}(\xtrue) = \mode( P(r_{E_\calib} \; | \; \xtrue))$.
In practice, the individual responses
$r_{E_\calib}$ are collected in histograms for each bin of $x_\true$ (where typically
$x_\true=(p_{\textrm{T}}^\true,m^\true,\eta^\true)$). From these
histograms, $R_{E_\calib}(x_\true)$ is extracted from a fit to a Gaussian
distribution\footnote{Firstly, the histograms are smoothed with a Savitzky--Golay filter~\cite{Savitzky1964} to improve the fit stability in low-statistics bins.
Then, heuristics are employed to limit the fit around
the central value by setting the fitting range around the mean of the histogram to be between $\pm N\sigma$, where $\sigma$ is the standard deviation.} and the calibrated response resolution
$\sigma$ from the calculation of the IQR divided by $R_{E_\calib}$ (not to be confused with $\sigma_\pred$ predicted by the network and described in Sec. \ref{subsubsec:loss}).
 
In the following, the performance is presented by plotting the
response or resolution as a function of one of the bin-defining
variables. Better performance is indicated by a response closer to
1.0 (called response closure) and a resolution closer to 0.0 in all of the phase space.
All plots include statistical uncertainties computed from the fits of the response distributions
but they are generally too small to be visible.

\subsubsection{Performance on QCD dijets}
 
\begin{figure}[htbp]
\centering
\subfloat[\centering]{{\includegraphics[width=7.55cm]{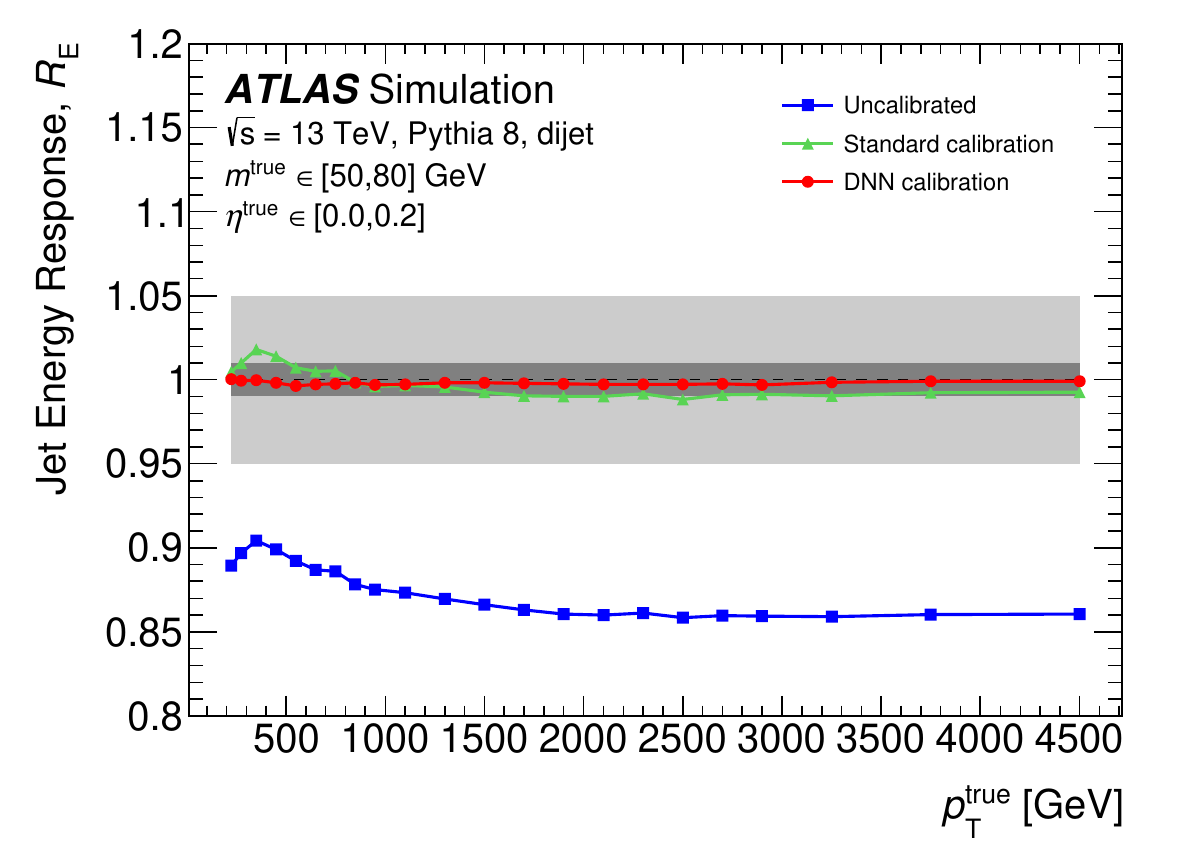} }}
\qquad
\subfloat[\centering]{{\includegraphics[width=7.55cm]{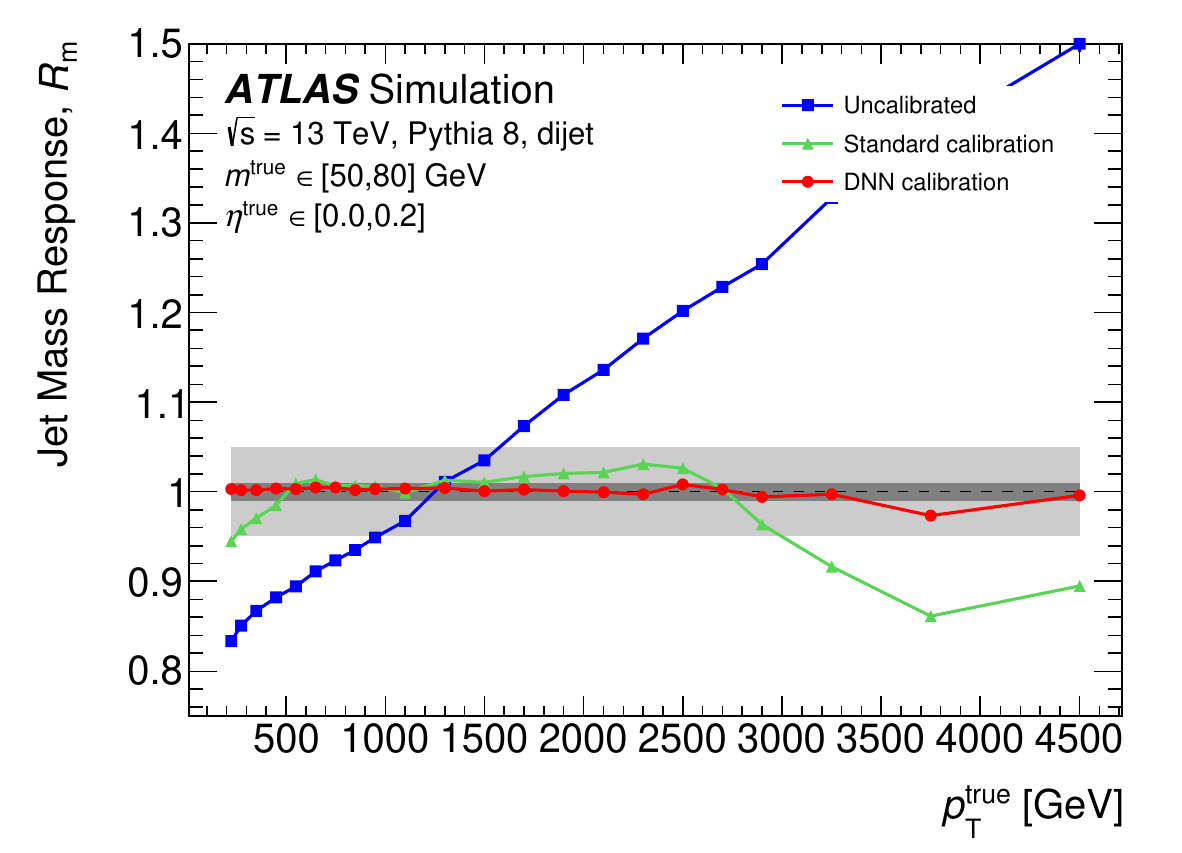} }}
\qquad
\subfloat[\centering]{{\includegraphics[width=7.55cm]{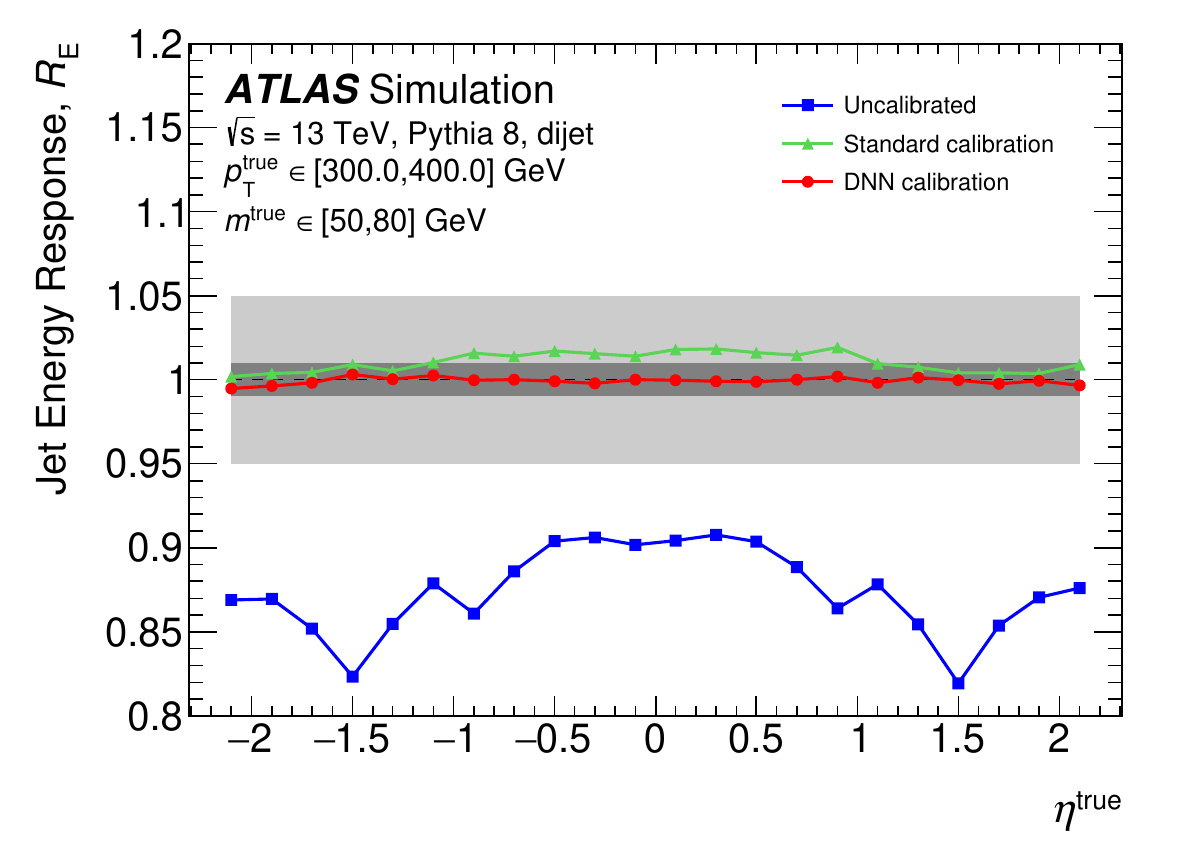} }}
\qquad
\subfloat[\centering]{{\includegraphics[width=7.55cm]{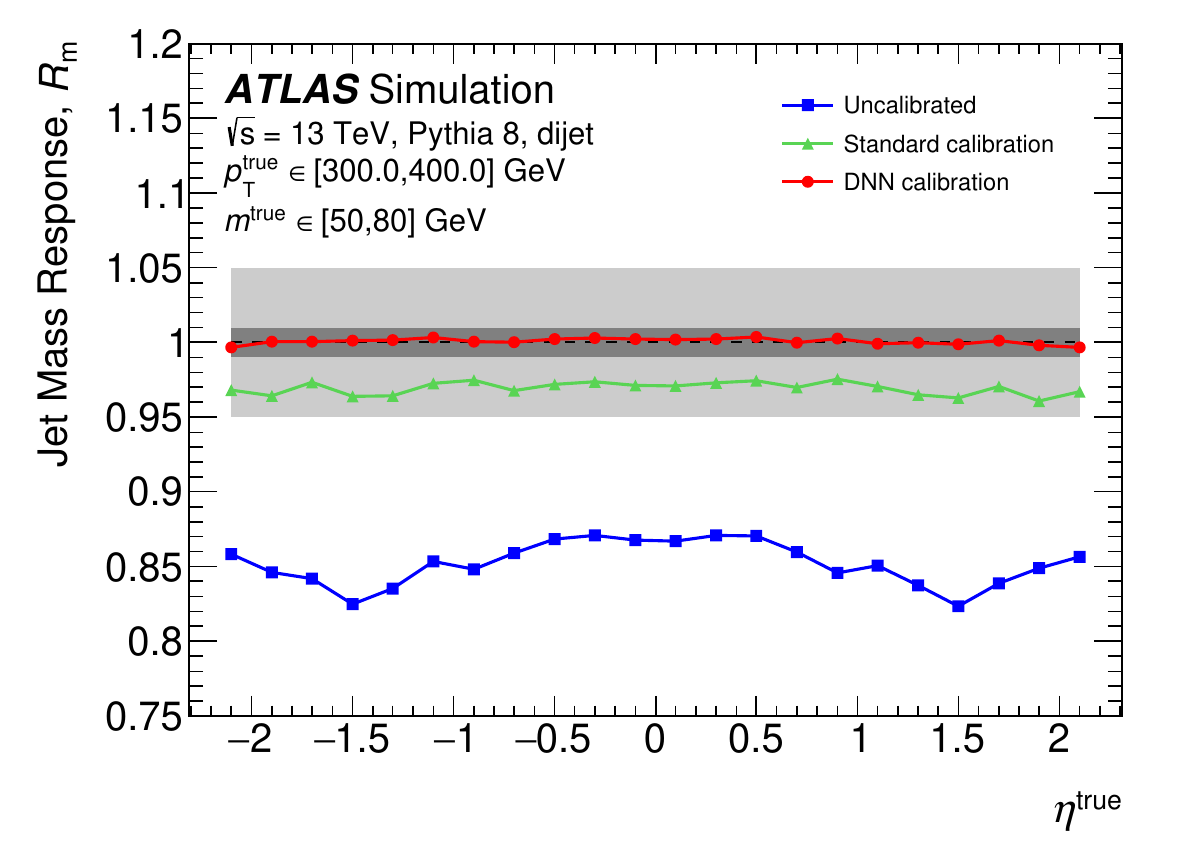} }}
\qquad
\subfloat[\centering]{{\includegraphics[width=7.55cm]{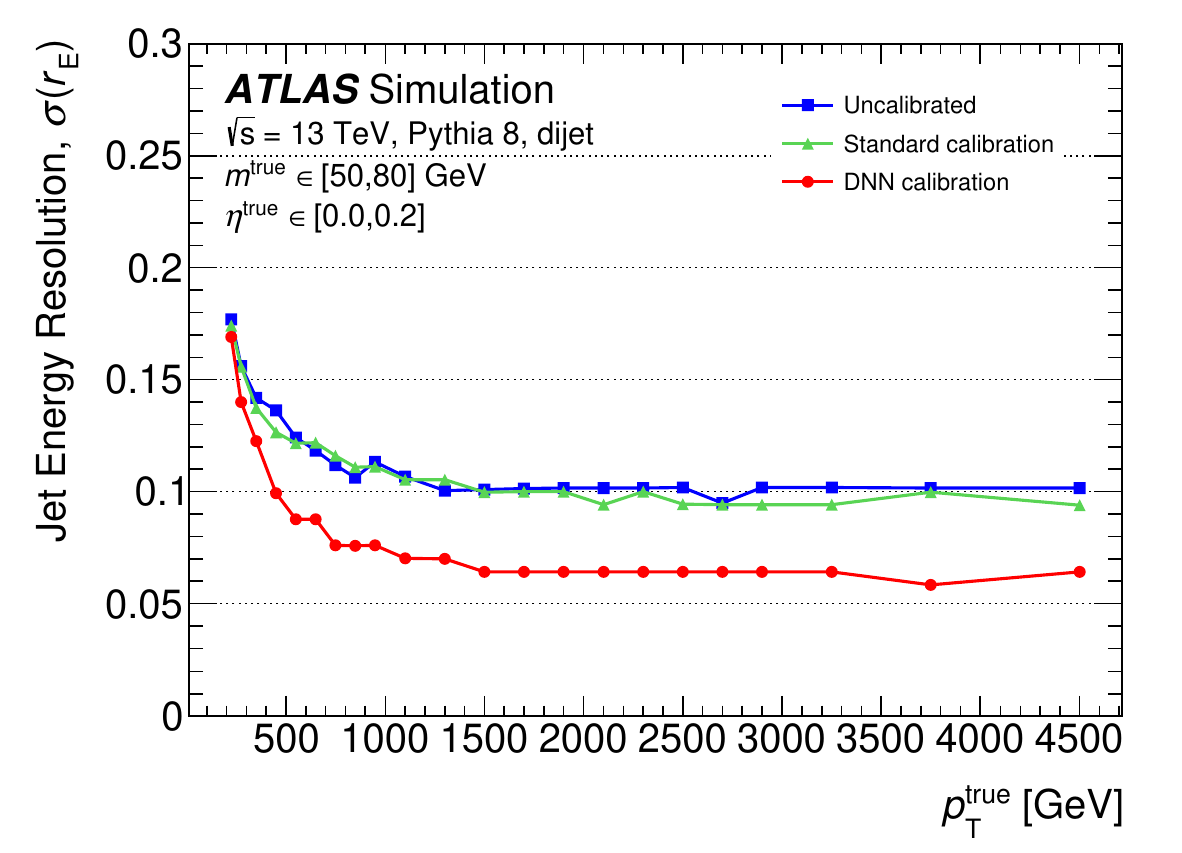} }}
\qquad
\subfloat[\centering]{{\includegraphics[width=7.55cm]{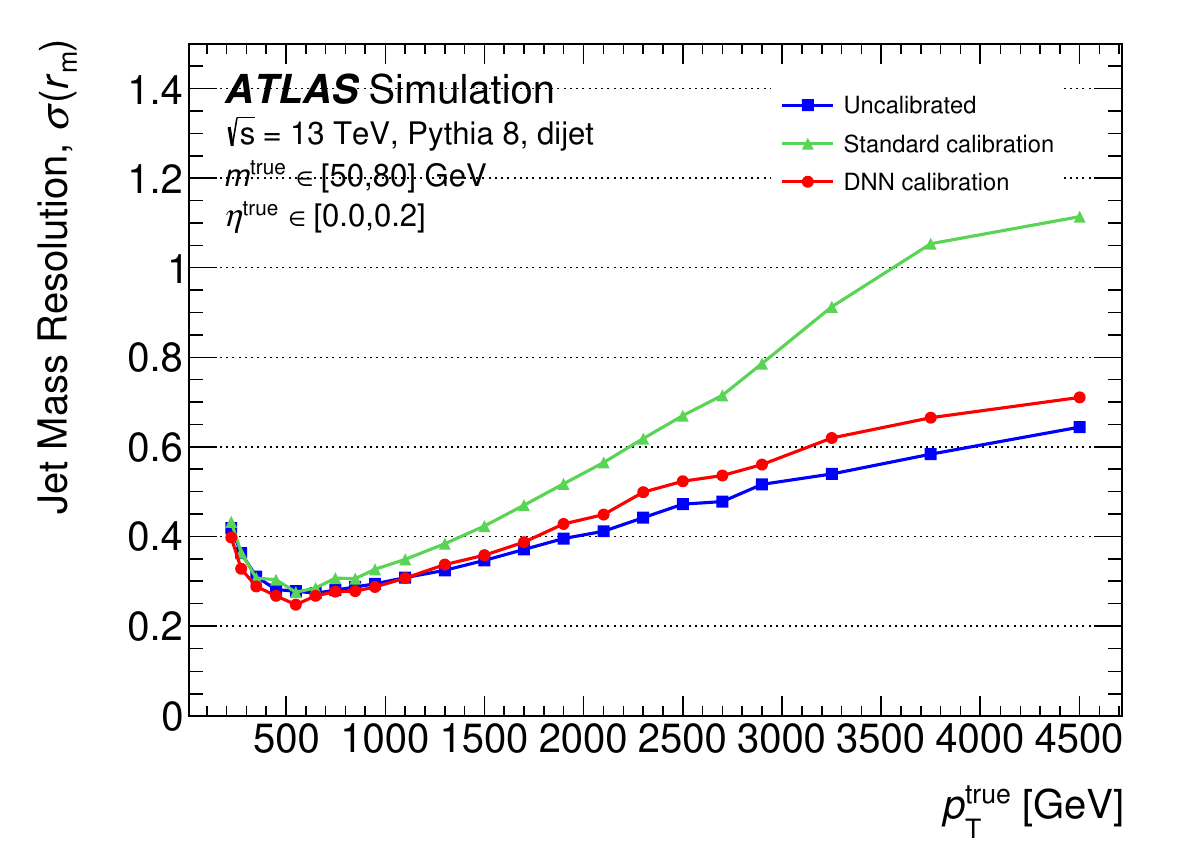} }}
\caption{Jet energy (left) and mass (right) responses (a, b, c, d) and resolutions (e, f) for different calibrations (squares: no calibration, triangles: standard calibration, circles: DNN calibration) as a function of $\pT^\true$ (a, b, e, f) and $\eta^\true$ (c, d). The horizontal bands in (a, b, c, d) represent 1$\%$ (dark) and 5$\%$ (light) up and down divergences from closure.}
\label{fig:emResp_Reso_QCD}
\end{figure}
 
As a closure test, the DNN calibration is applied to the QCD sample
used during training. This allows verification that the DNN convergence translates into good physics performance in a sample with large statistics. Overall, the calibration is found to perform
very well for $|\eta|<2$ with the best results at high mass in terms
of response closure and resolution. Figure~\ref{fig:emResp_Reso_QCD}
presents the energy and mass calibrated responses as a function of $\pT^\true$ and
$\eta^\true$ and the energy and mass resolution as a function of $\pT^\true$ for specific bins.
It shows excellent response closure for both energy and mass, closer to unity than the standard
calibration~\cite{JETM-2018-05}. The DNN calibration also shows an important improvement in
the resolution relative to the standard calibration. For the mass,
the uncalibrated resolution appears better than the calibrated
ones. However, this is a relative quantity (as defined in Section
\ref{sec:std_calib}) while the absolute value of the IQR scales with the
central value. As a consequence, comparing the uncalibrated resolution corresponding to responses significantly
differing from 1.0 with a calibrated resolution is not very meaningful.
 
\begin{figure}[htbp]
\centering
\subfloat[\centering]{{\includegraphics[width=7.55cm]{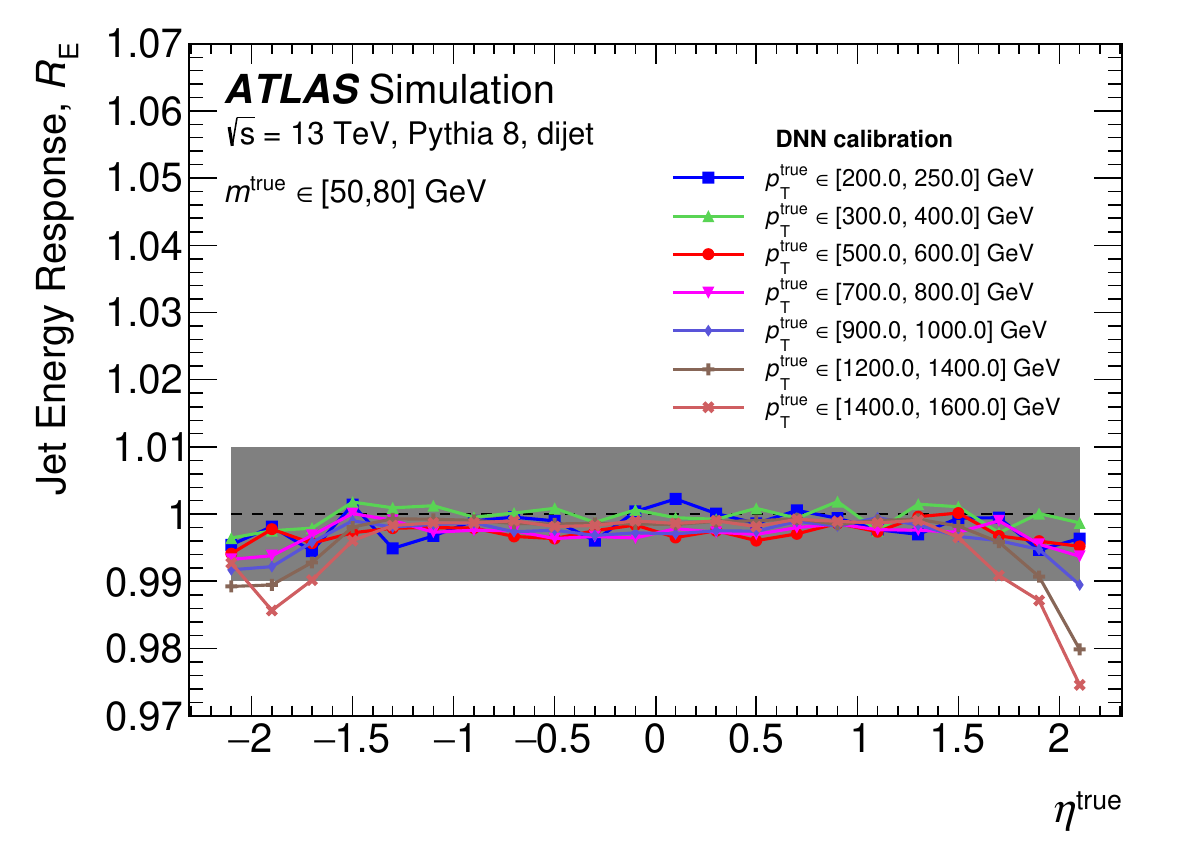} }}
\qquad
\subfloat[\centering]{{\includegraphics[width=7.55cm]{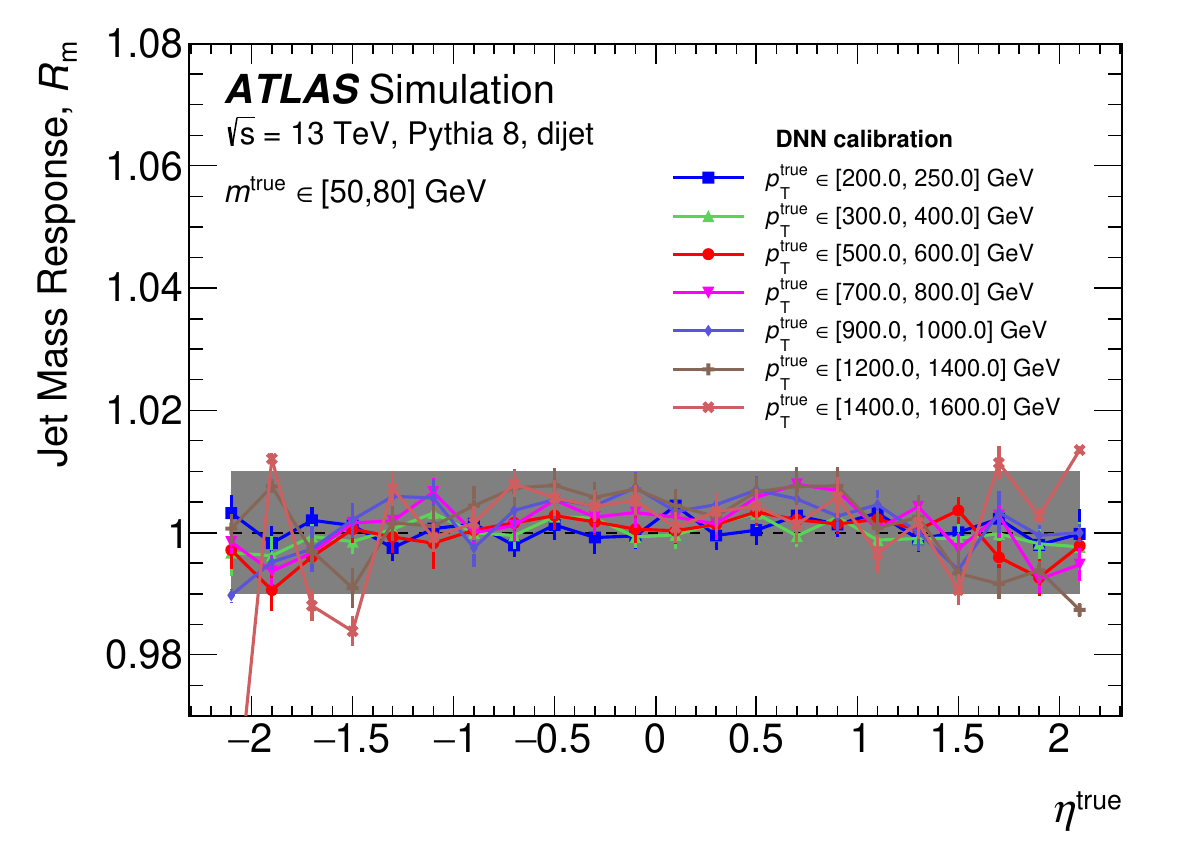} }}
\caption{The (a) jet energy and (b) mass calibrated response for the DNN calibration as a function of $\eta^\true$ and for multiple bins in $\pT^\true$. The horizontal bands represent 1$\%$ (dark) and 5$\%$ (light) up and down divergences from closure.}
\label{fig:emResp_DNNonly_QCD}
\end{figure}

Figure~\ref{fig:emResp_DNNonly_QCD} presents the excellent consistency of the DNN calibration performance for different $\pT^\true$ bins.
 
In every bin of $(\pT^\true, m^\true, \eta^\true)$ with
$\pT^\true>\qty{200}{\GeV}$, $m^\true>\qty{40}{\GeV}$  and
$|\eta^\true|<2$, the DNN calibration is equally or more performant
than the standard calibration with a better response closure and resolution.
Improvements of a few percent up to around $10\%$ are
observed in both energy and mass response closure and resolution.

\subsubsection{Effect of $\eta$ annotation}
 
\begin{figure}[htbp]
\centering
\subfloat[\centering]{{\includegraphics[width=7.55cm]{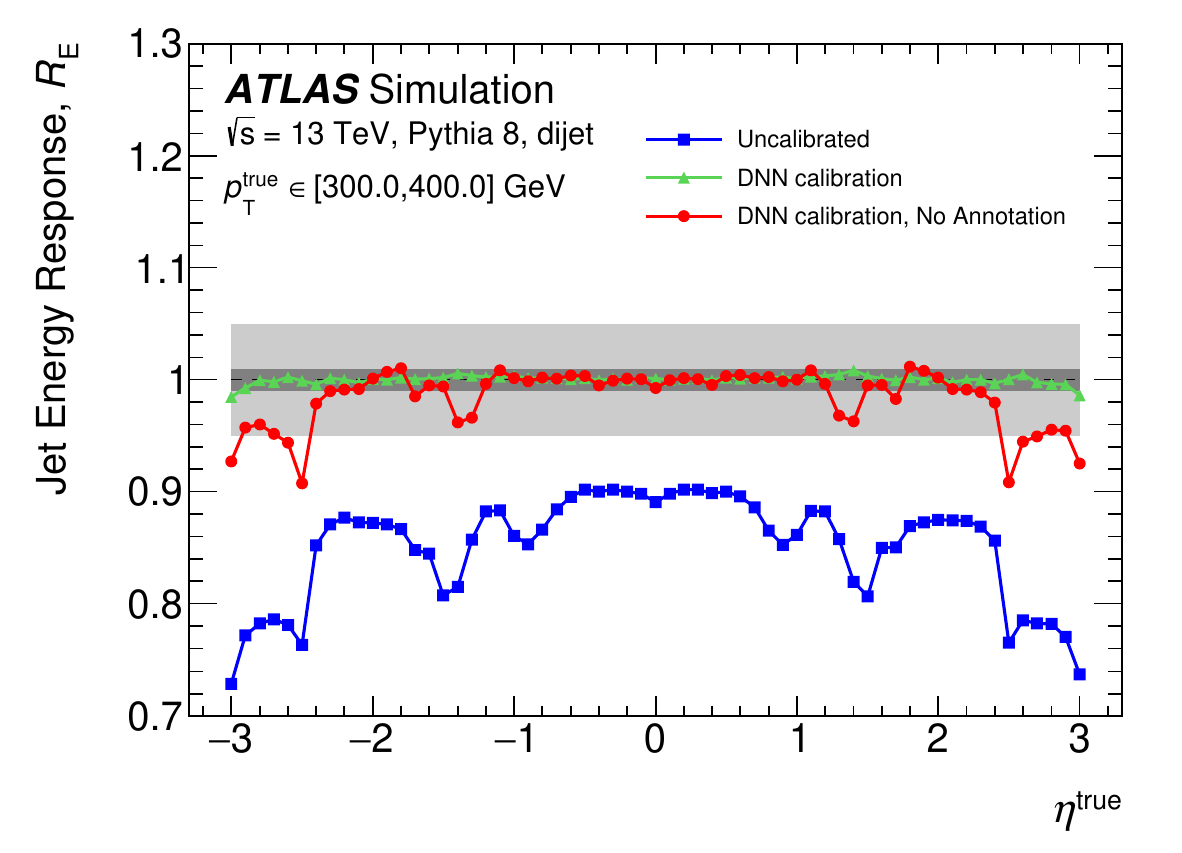} }}
\qquad
\subfloat[\centering]{{\includegraphics[width=7.55cm]{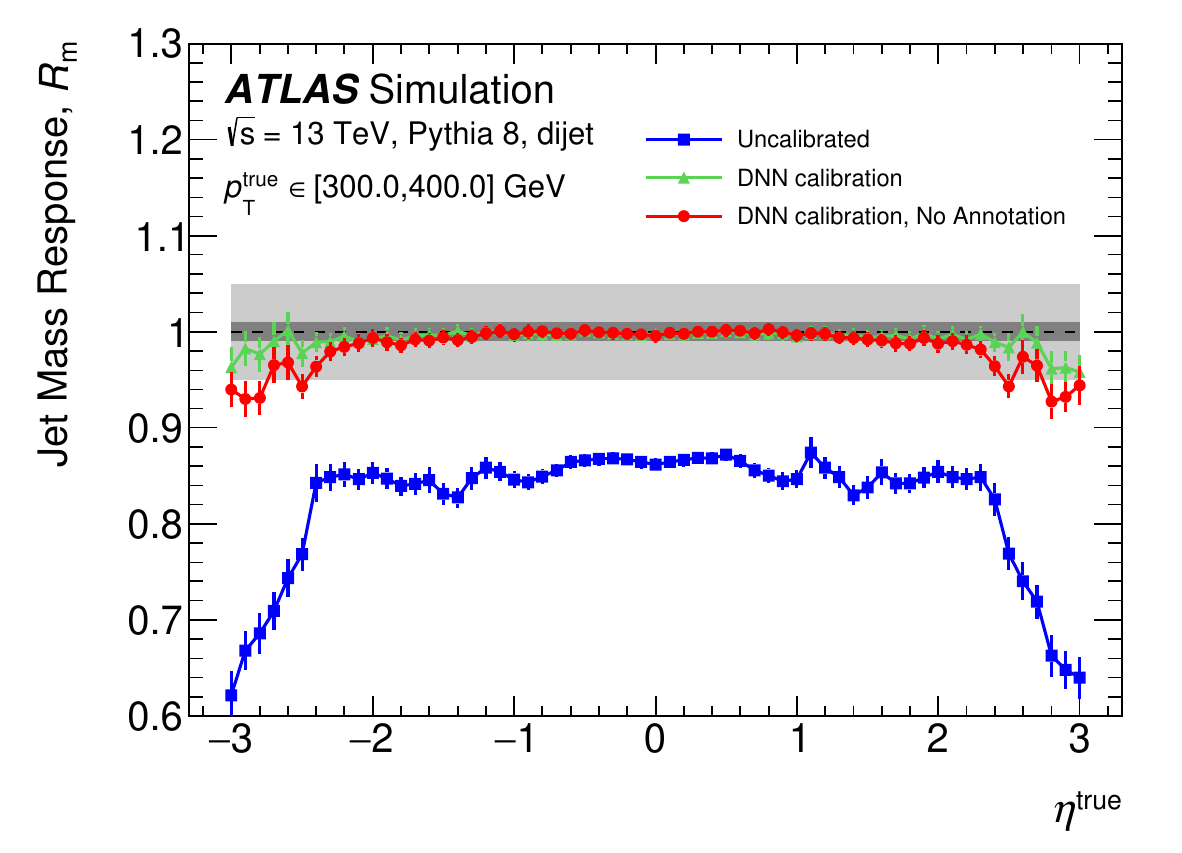} }}
\caption{The (a) jet energy and (b) mass response before calibration (squares) and after the DNN calibration with (triangles) or without (circles) the $\eta$ annotation step as a function of $\eta^\true$. The horizontal bands represent 1$\%$ (dark) and 5$\%$ (light) up and down divergences from closure.}
\label{fig:emResp_NoAnnot_QCD}
\end{figure}

Section~\ref{subsubsec:features} describes the necessity to add the
$\eta$ annotation to help the DNN
adapt to the complex detector geometry. The assumption is that the extra information
represented by the angular encoding together with the related weights allows the training process
to specialise the predictions according to the detector region.
As a validation step, Figure~\ref{fig:emResp_NoAnnot_QCD}
shows a comparison of the jet energy and mass
response closure between two DNN setups with the same training
procedure  and the same architecture except
for the $\eta$ annotation step. Its inclusion clearly improves the
DNN calibrated response closure in the almost complete $\eta$ range and especially in the calorimeter
transition region above $|\eta|=1.2$. The JMS improvement is less
important, as expected since the uncalibrated response (squared
markers) variations for the mass are less important than for the
energy.

These improvements are present in every \pT and $\eta$ bin, proving the importance
of the $\eta$ annotation step.

\subsubsection{Performance on other topologies}

Physics analyses mainly use \largeR jets to reconstruct the decays of
massive and boosted particles like top quarks, $W$, $Z$ or Higgs bosons.
These decays result in jets with different shower
topologies than those originating from the dijet samples with which the DNN
is trained; this can have an impact on the mass response as demonstrated
in Ref.~\cite{JETM-2022-01} where significant differences between $W$-initiated and
$q/g$-initiated jets were observed. It is hence very important to
verify that the DNN predictions generalise to other such topologies as
the same calibration is applied to all jets in data as their origin will be unknown.
For this, simulated samples of
processes with heavy hadronically decaying particles, as described
in Section~\ref{subsec:montecarlo}, are considered. For each sample, the same jet
selection and DNN calibration as for QCD jets are applied and the
resulting scale and resolution are evaluated in mass bins centred on
the mass of the corresponding massive particle. The population of these massive,
high-momentum jets is then expected to be entirely dominated by the
hadronic decays of the heavy particles considered so that the effect of the topology can be assessed.
 
\begin{figure}[htbp]
\centering
\subfloat[\centering]{{\includegraphics[width=7.55cm]{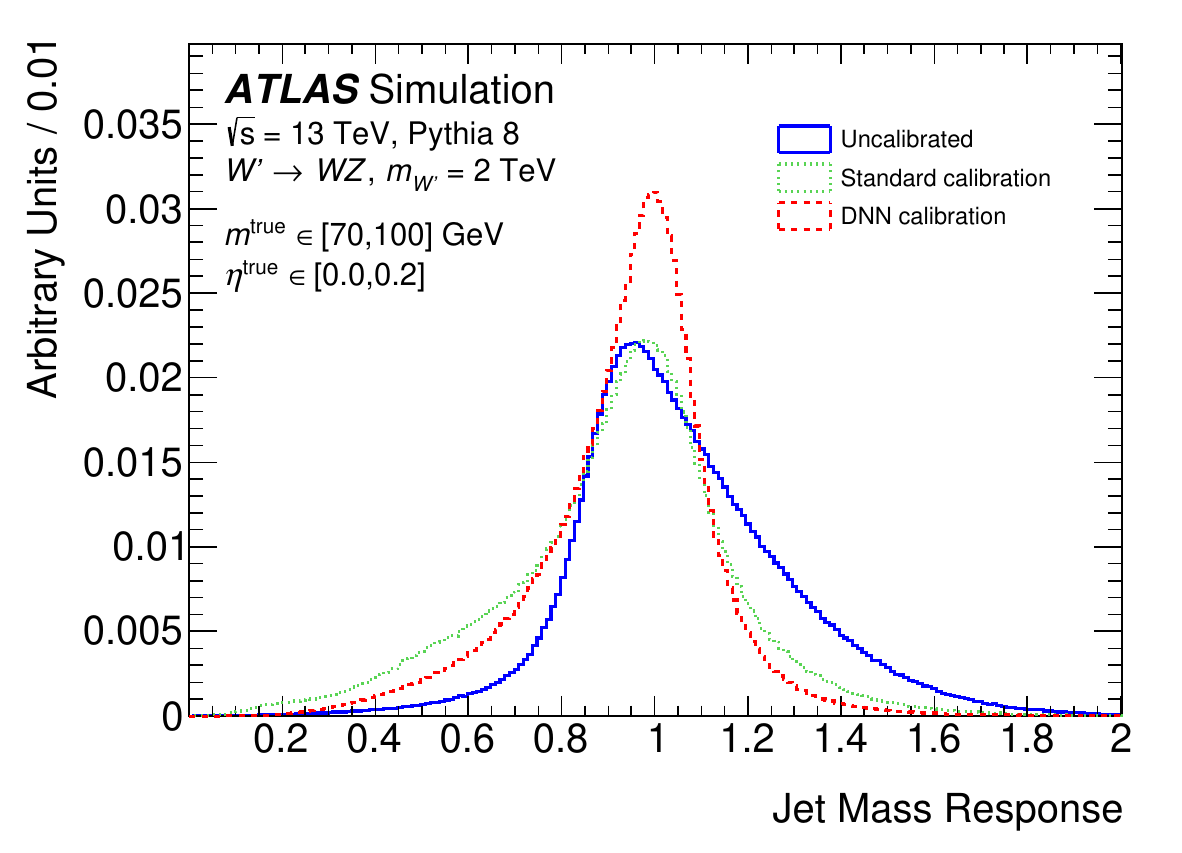} }}
\qquad
\subfloat[\centering]{{\includegraphics[width=7.55cm]{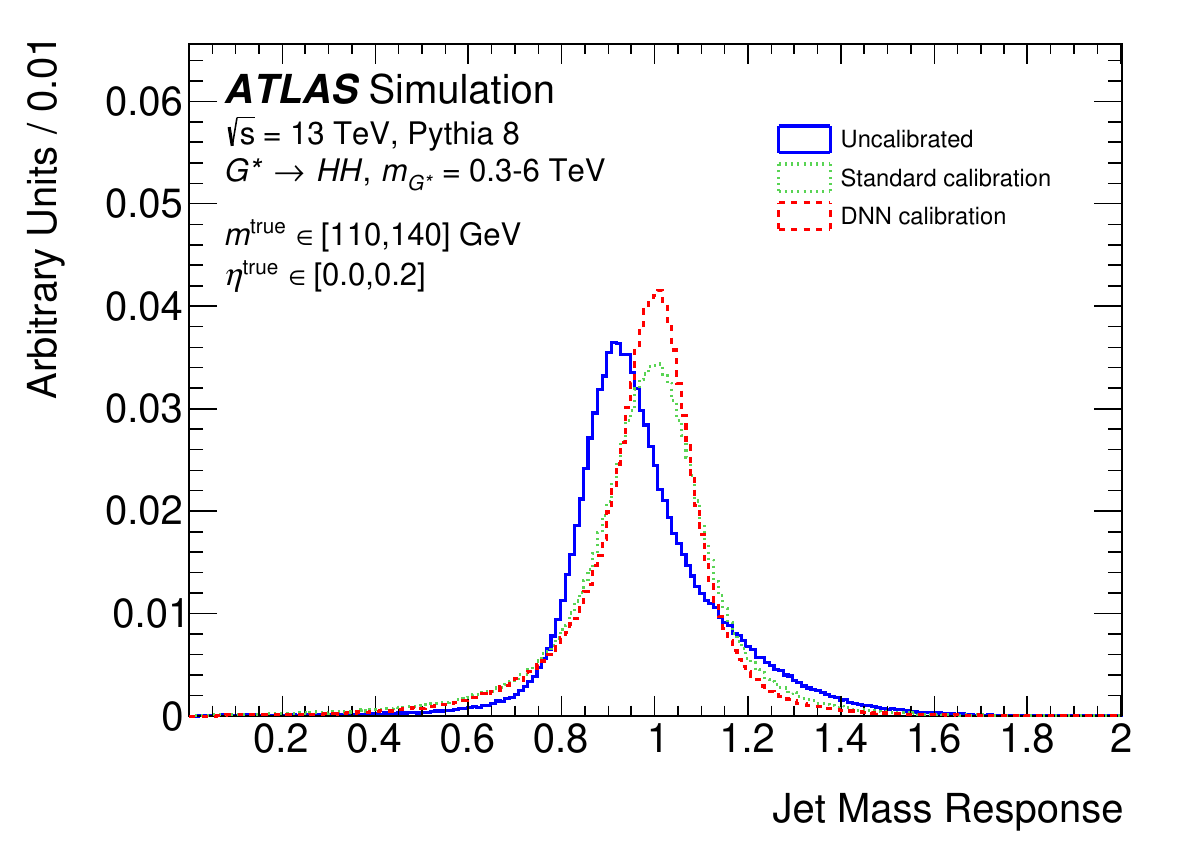} }}
\qquad
\subfloat[\centering]{{\includegraphics[width=7.55cm]{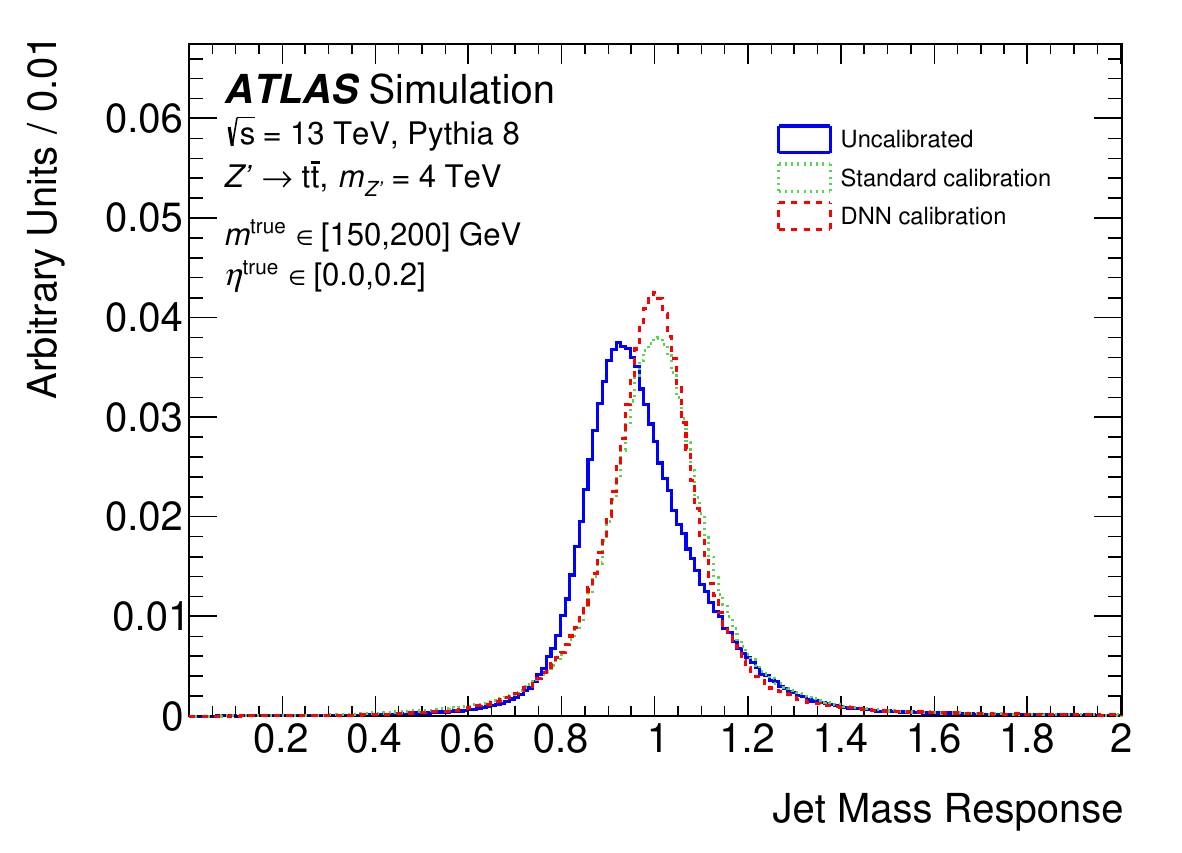} }}
\caption{(a) Boosted $W/Z$, (b) Higgs boson and (c) top-quark mass response distributions (normalised to unity) for different calibrations: uncalibrated (plain line), standard calibration (dotted line) and DNN calibration (dashed line).}
\label{fig:mR_hist_boosted}
\end{figure}

Figure~\ref{fig:mR_hist_boosted} presents the jet mass response
distributions obtained with boosted massive jets ($W/Z$, Higgs boson and top-quark
decay) for different calibrations (uncalibrated, standard calibration
and DNN calibration). It shows the benefits of the DNN calibration
compared with the standard calibration or the absence of a calibration;
the distributions are narrower with less asymmetric tails and are better centered
on 1.0.
 
\begin{figure}[htbp]
\centering
\subfloat[\centering]{{\includegraphics[width=7.55cm]{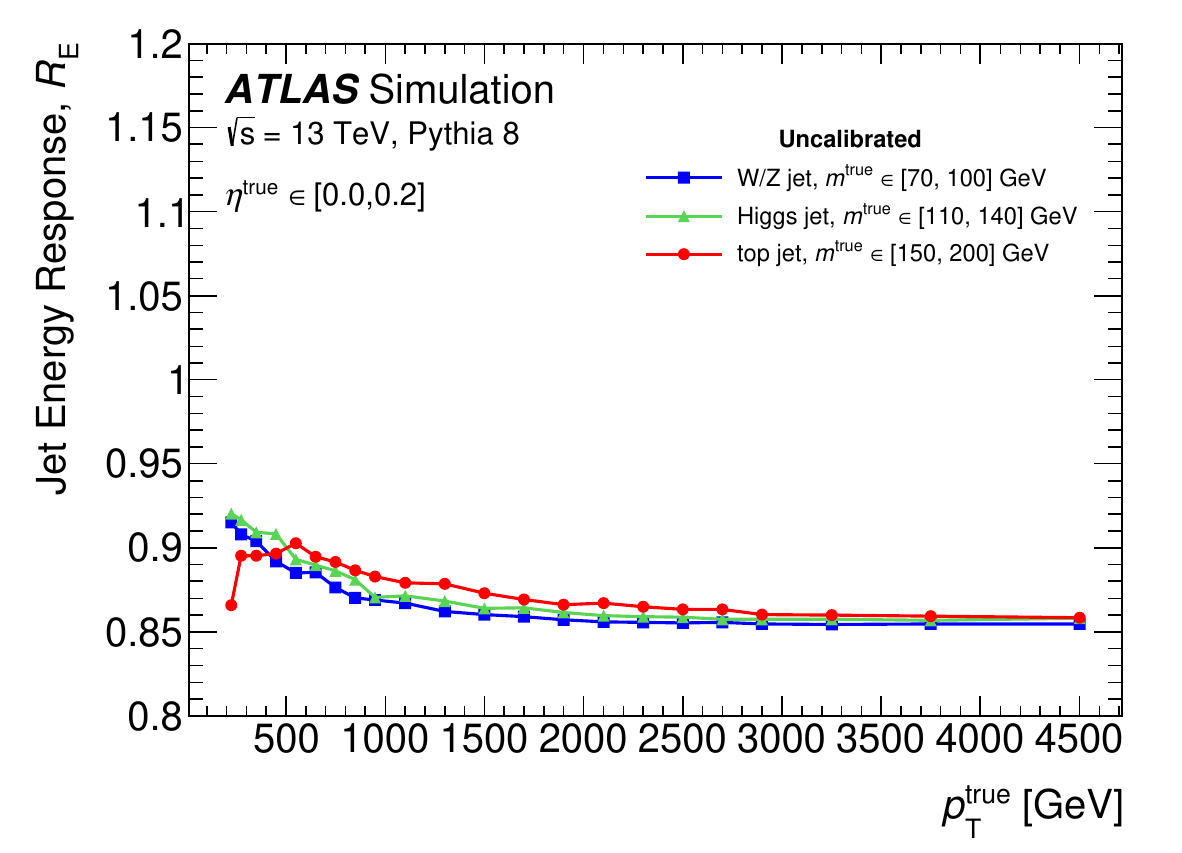} }}
\qquad
\subfloat[\centering]{{\includegraphics[width=7.55cm]{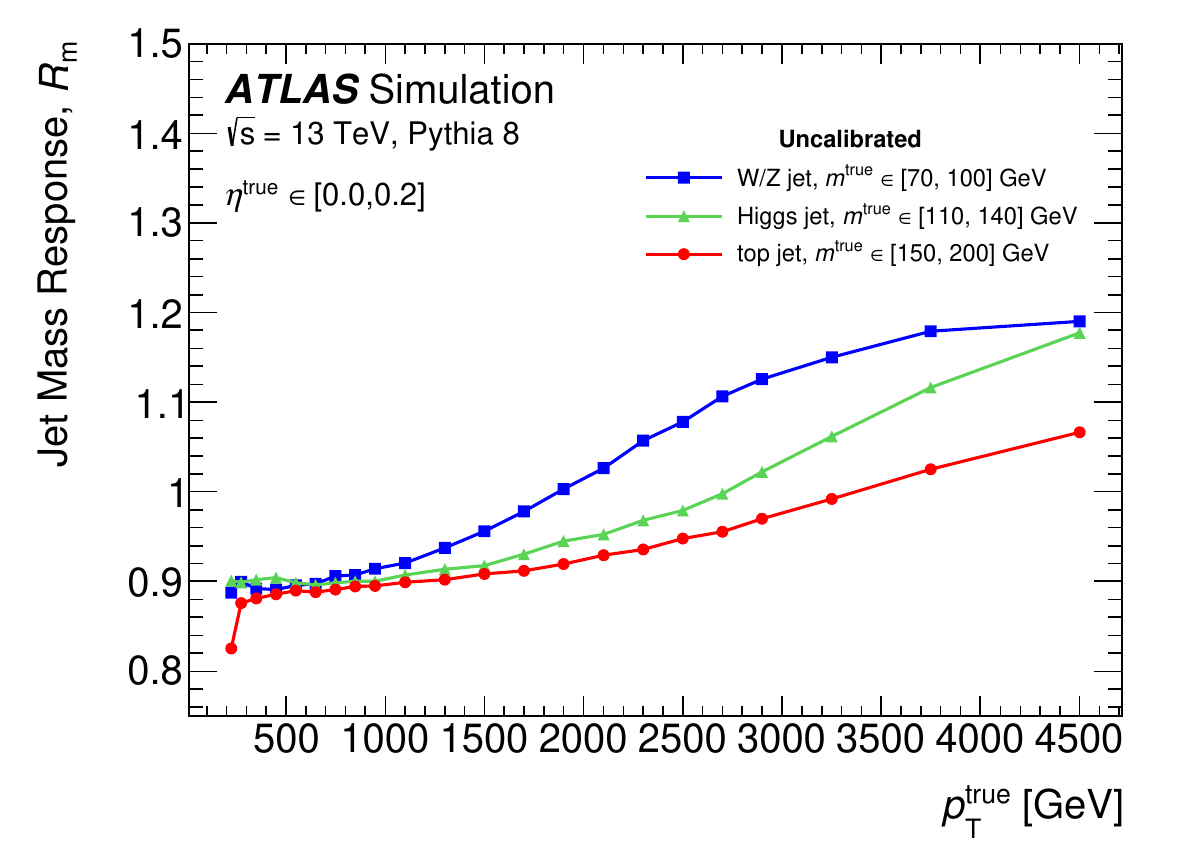} }}
\qquad
\subfloat[\centering]{{\includegraphics[width=7.55cm]{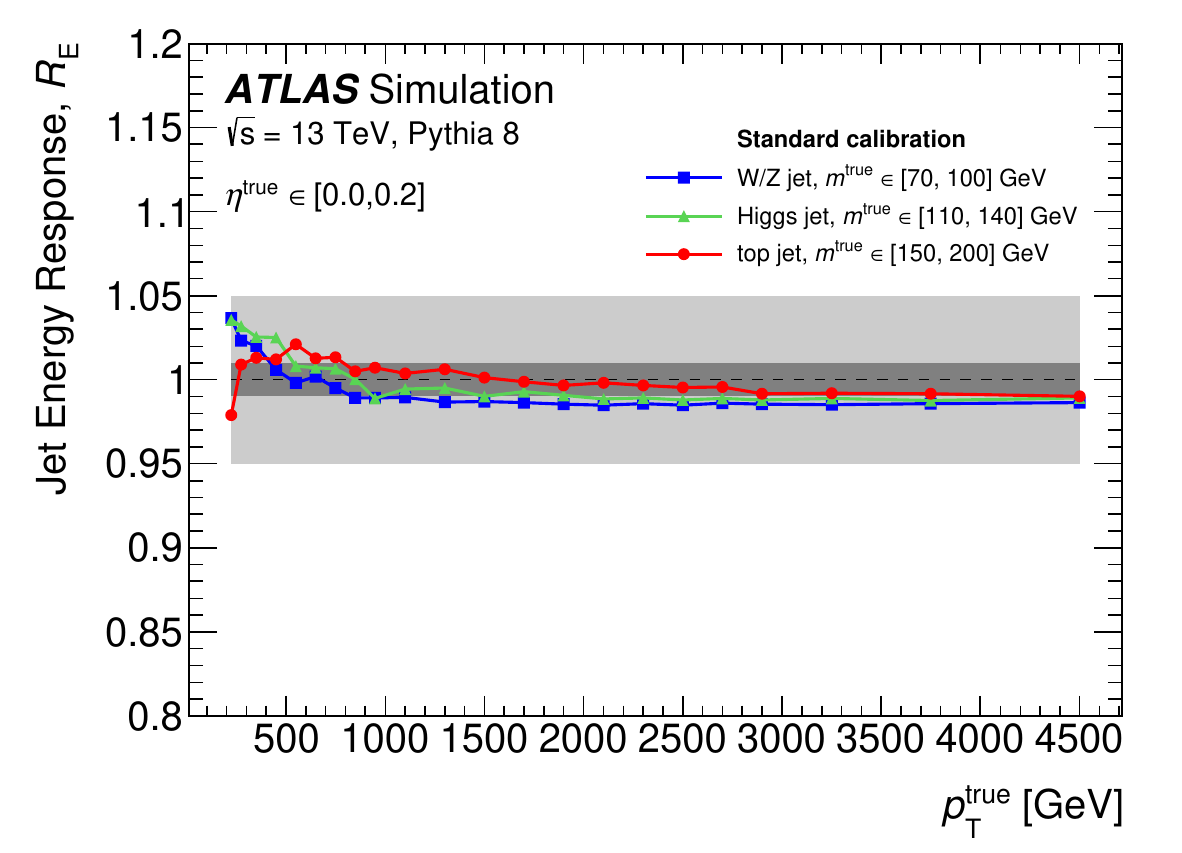} }}
\qquad
\subfloat[\centering]{{\includegraphics[width=7.55cm]{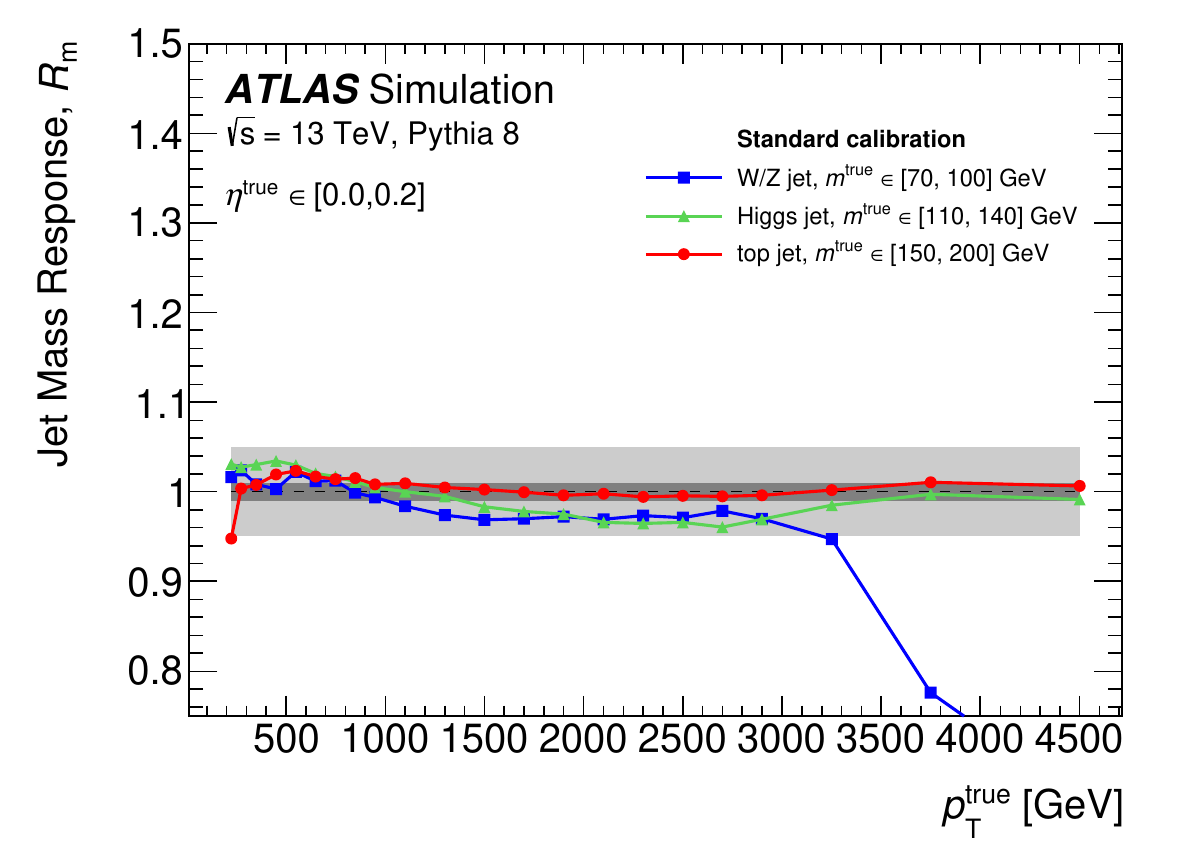} }}
\qquad
\subfloat[\centering]{{\includegraphics[width=7.55cm]{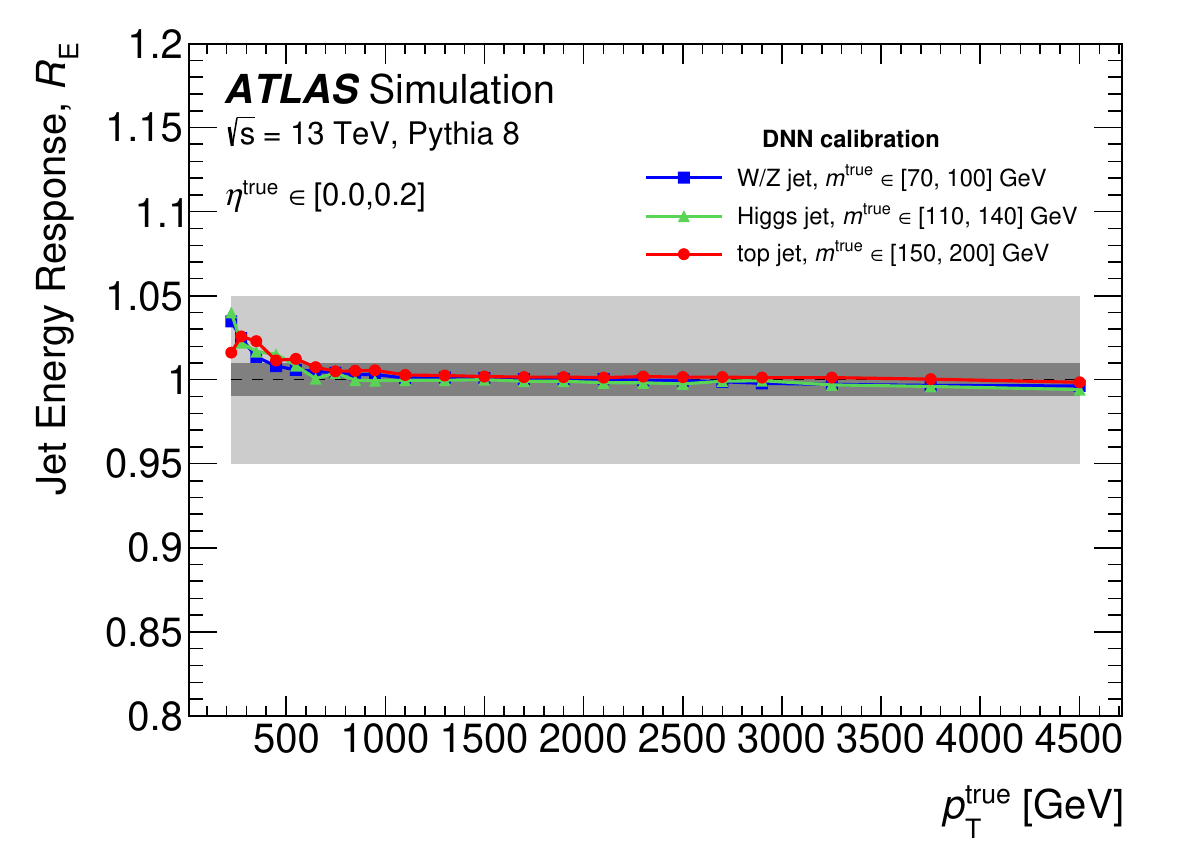} }}
\qquad
\subfloat[\centering]{{\includegraphics[width=7.55cm]{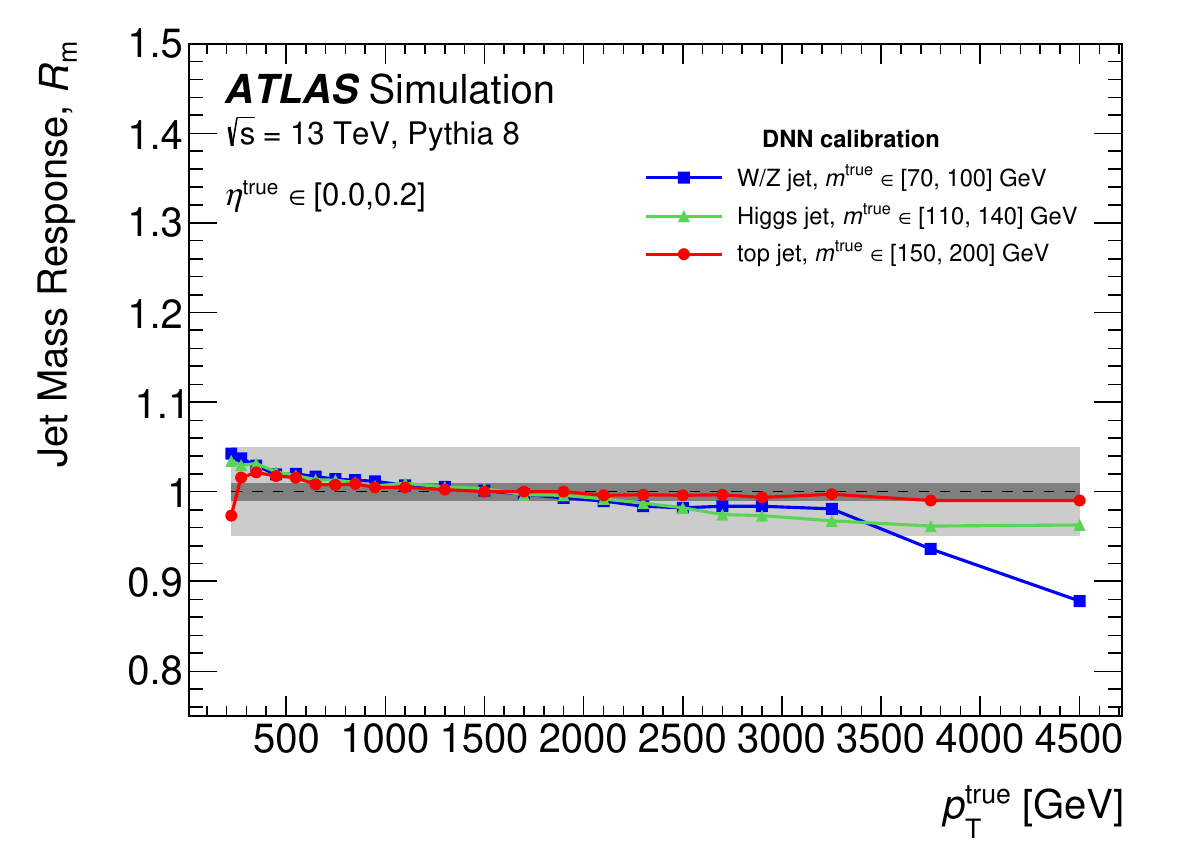} }}
\caption{Boosted W/Z, top-quark and Higgs boson jet energy (left) and mass (right) responses as a function of $\pT^\true$ for different calibrations: (a, b) no calibration, (c, d) standard calibration and (e, f) DNN calibration. The horizontal bands represent 1$\%$ (dark) and 5$\%$ (light) up and down divergences from closure.}
\label{fig:emResp_boosted}
\end{figure}

Figure \ref{fig:emResp_boosted} shows the jet energy and mass average
responses as a function of the true \pT for the same samples and calibrations.
As expected, large differences between the different topologies are
observed before calibration, especially in the mass response. While
both the standard and the DNN calibrations reduce these response
differences, the DNN calibration outperforms the standard one as it
gives better closure for each topology, showing that the DNN
predictions generalise well to more complex topologies. A small non-closure in the energy response is however observed for all the boosted topologies at low $\pT^\true$ for the
DNN calibration. For the $W/Z$ and Higgs boson jets, this small under-performance can easily be explained by the internal
structure of the boosted massive jets; at high momentum, $W/Z$ and Higgs boson \largeR jets are usually composed of two sub-jets
with an angular separation of $\Delta R\sim \frac{2m}{\pT}$. At low \pT, their angular separation becomes comparable
to the \largeR jet radius of 1.0, and the structure becomes neither 1-prong nor 2-prong. Such a structure is
possibly rare in the dijet training sample or not captured by the input substructure variables, making it
impossible for the DNN to learn to correct for this effect. For top-quark jets, a similar effect
could be at play when the 3-prong decay $t\to Wb\to qq'b$ is not fully
contained within the jet radius, thus leading to atypical jets.
To solve this difficulty, a dedicated additional network structure and
training can be envisaged and is discussed in
Section~\ref{sec:prospects}.
 
\begin{figure}[htbp]
\centering
\subfloat[\centering]{{\includegraphics[width=7.55cm]{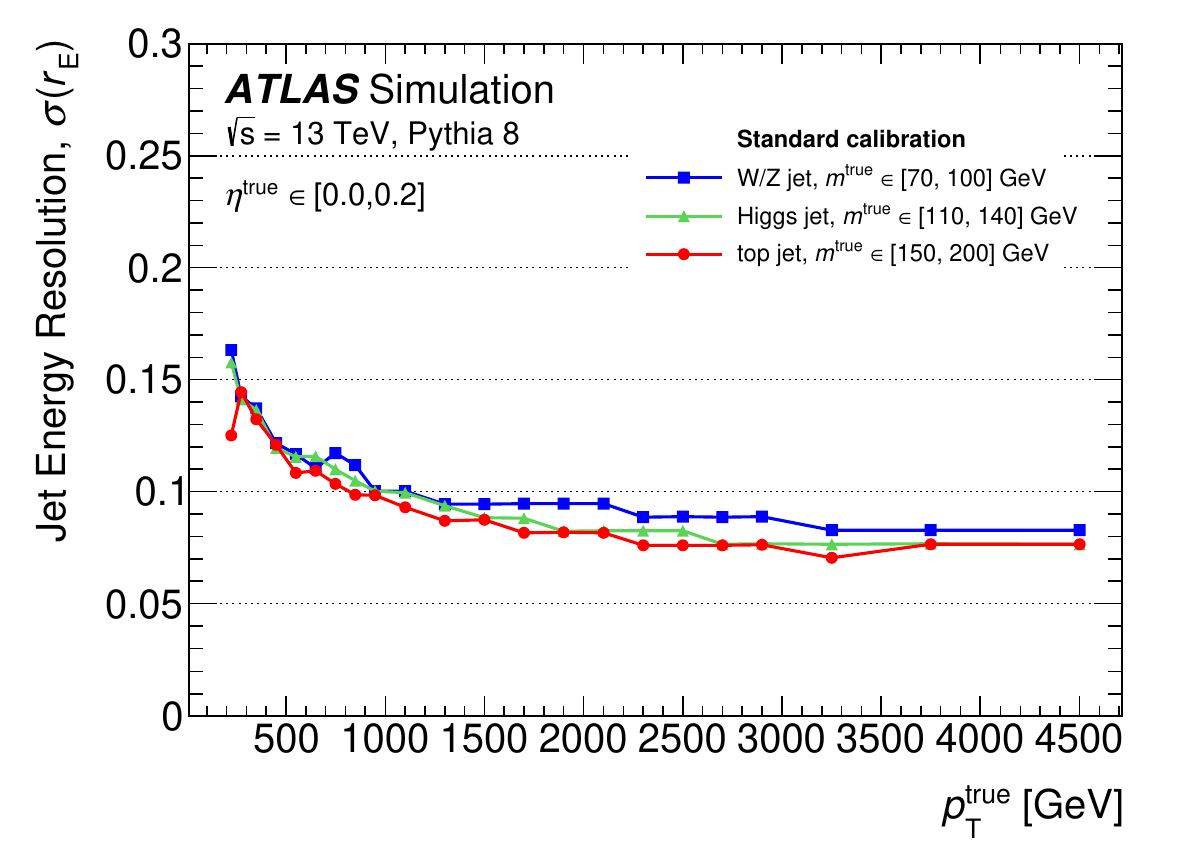} }}
\qquad
\subfloat[\centering]{{\includegraphics[width=7.55cm]{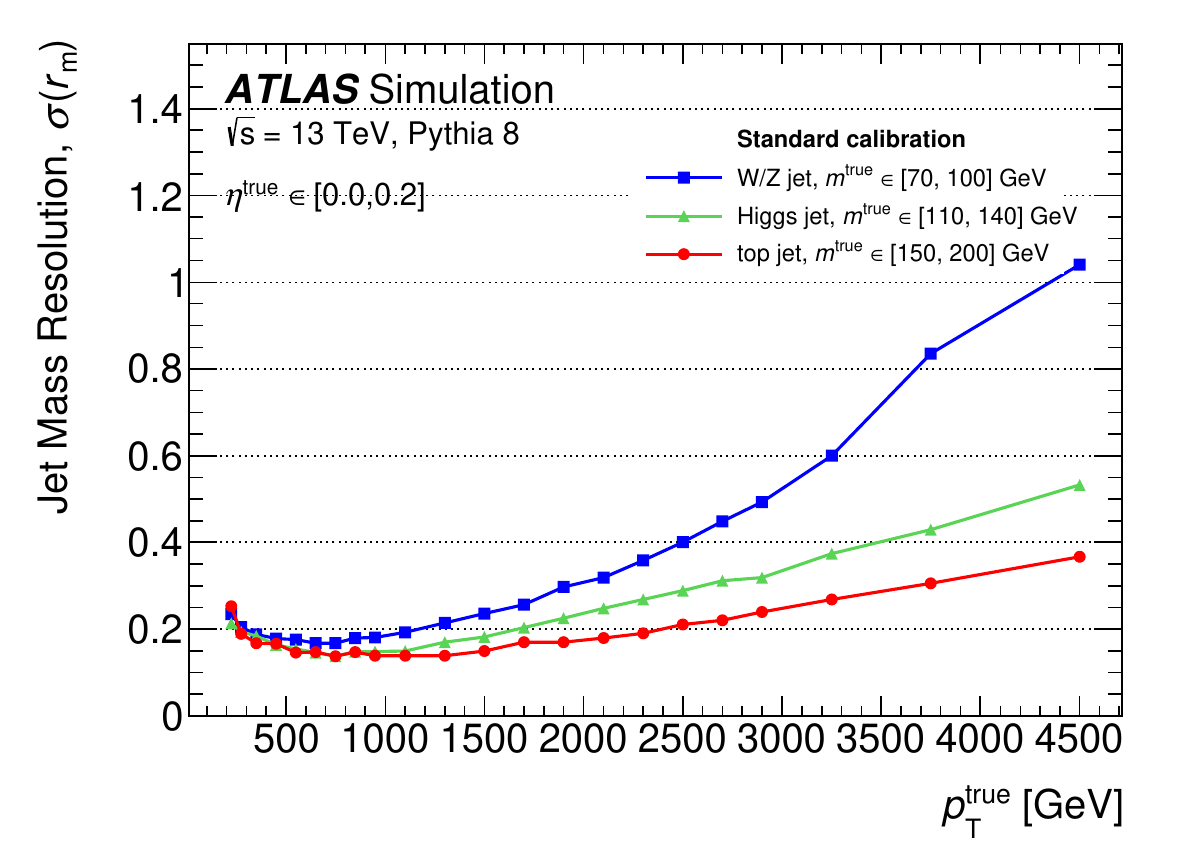} }}
\qquad
\subfloat[\centering]{{\includegraphics[width=7.55cm]{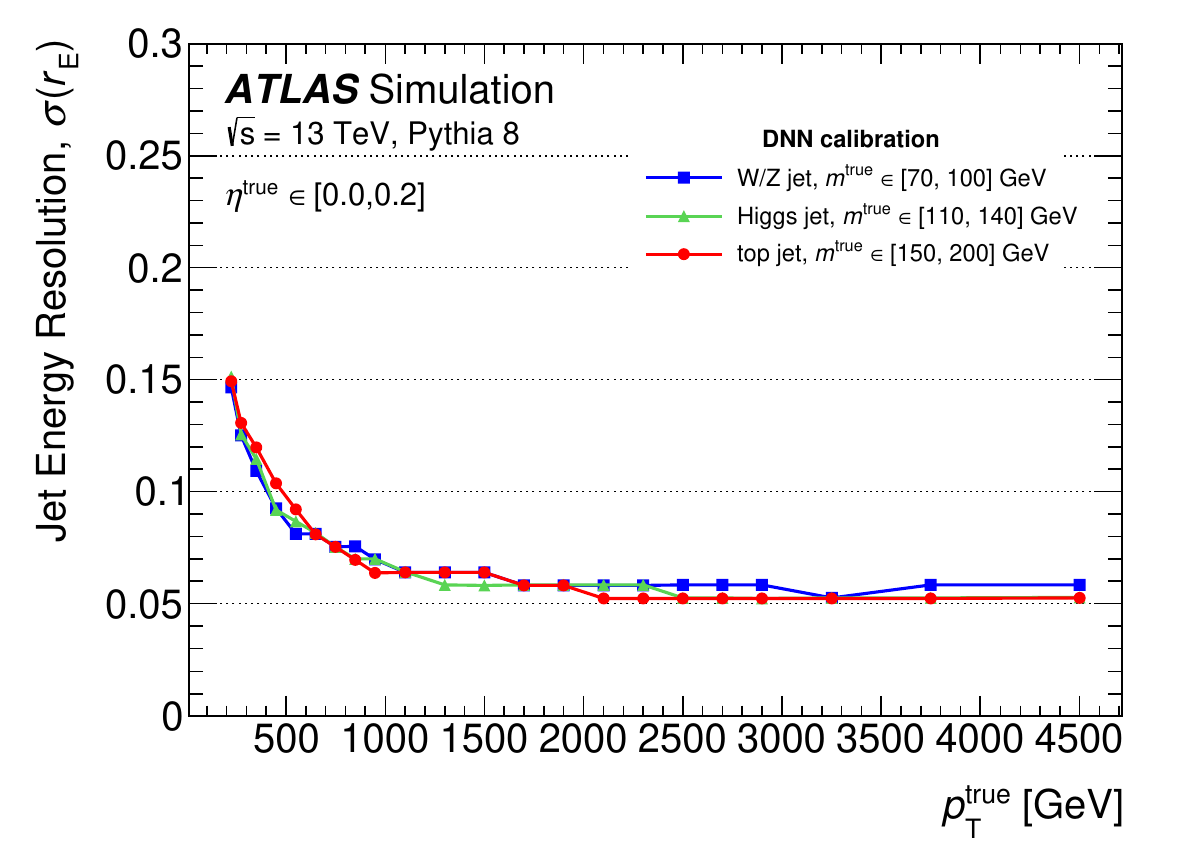} }}
\qquad
\subfloat[\centering]{{\includegraphics[width=7.55cm]{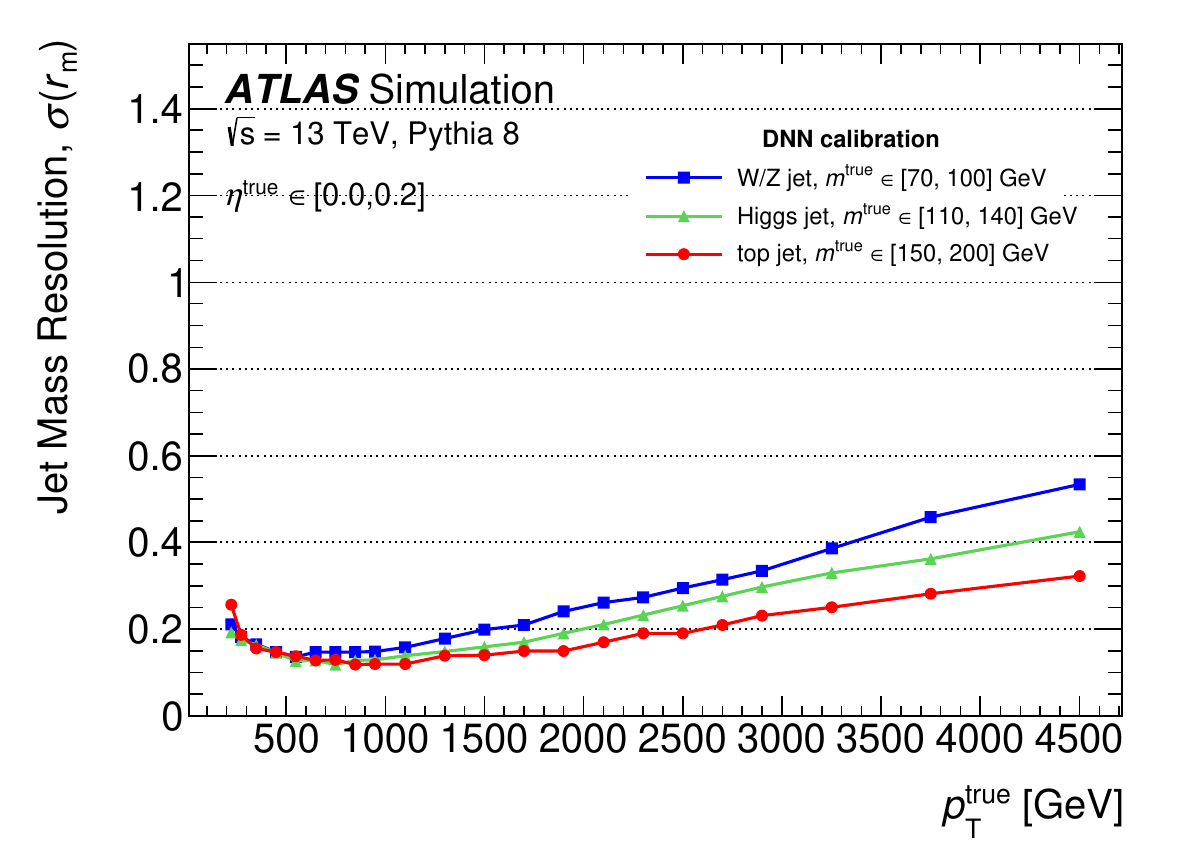} }}
\caption{Boosted $W/Z$, top-quark and Higgs boson jet energy (left) and mass
(right) resolutions as a function of $\pT^\true$ for different
calibrations: (a, b) standard calibration and (c, d) DNN calibration.}
\label{fig:emResolution_boosted}
\end{figure}
 
Figure \ref{fig:emResolution_boosted} presents the jet energy and mass
resolution for the boosted topologies. Here as well, the DNN
calibration shows improved performance relative to the standard
calibration. The resolution differences between the topologies here can be
explained by the increasing mass bins in which the responses are
evaluated; the resolution usually gets better at higher masses.

\subsection{Validation of the DNN calibration}
\label{subsec:validation}
 
An extensive series of tests are performed to ensure the good physical
behaviour of the DNN calibration. As described above, both the standard and DNN
calibrations are tested and compared.

\paragraph{Spectrum dependency:}
\label{subsubsec:spectrumdep}
 
\begin{figure}[htbp]
\centering
\subfloat[\centering]{\label{fig:emResp_LogEW_QCD_dist} {\includegraphics[width=7.55cm]{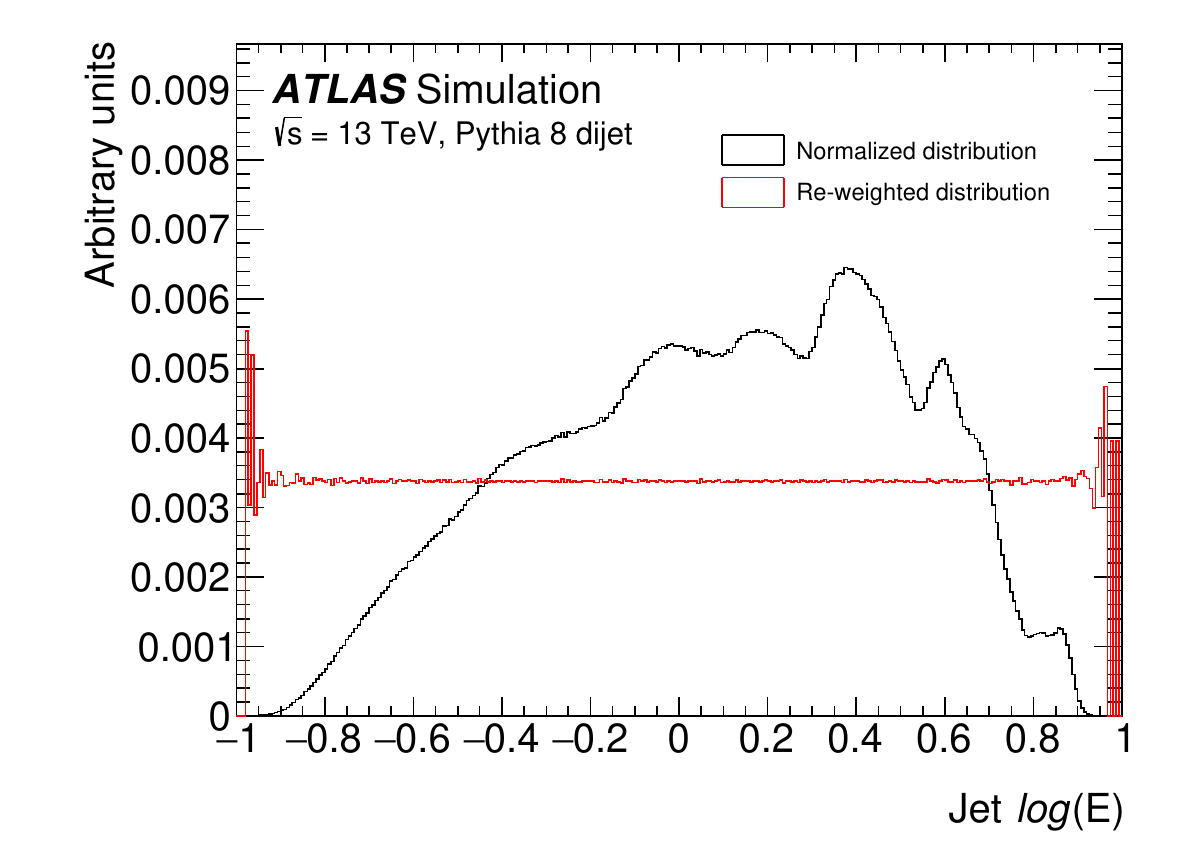} }}
\qquad
\subfloat[\centering]{{\includegraphics[width=7.55cm]{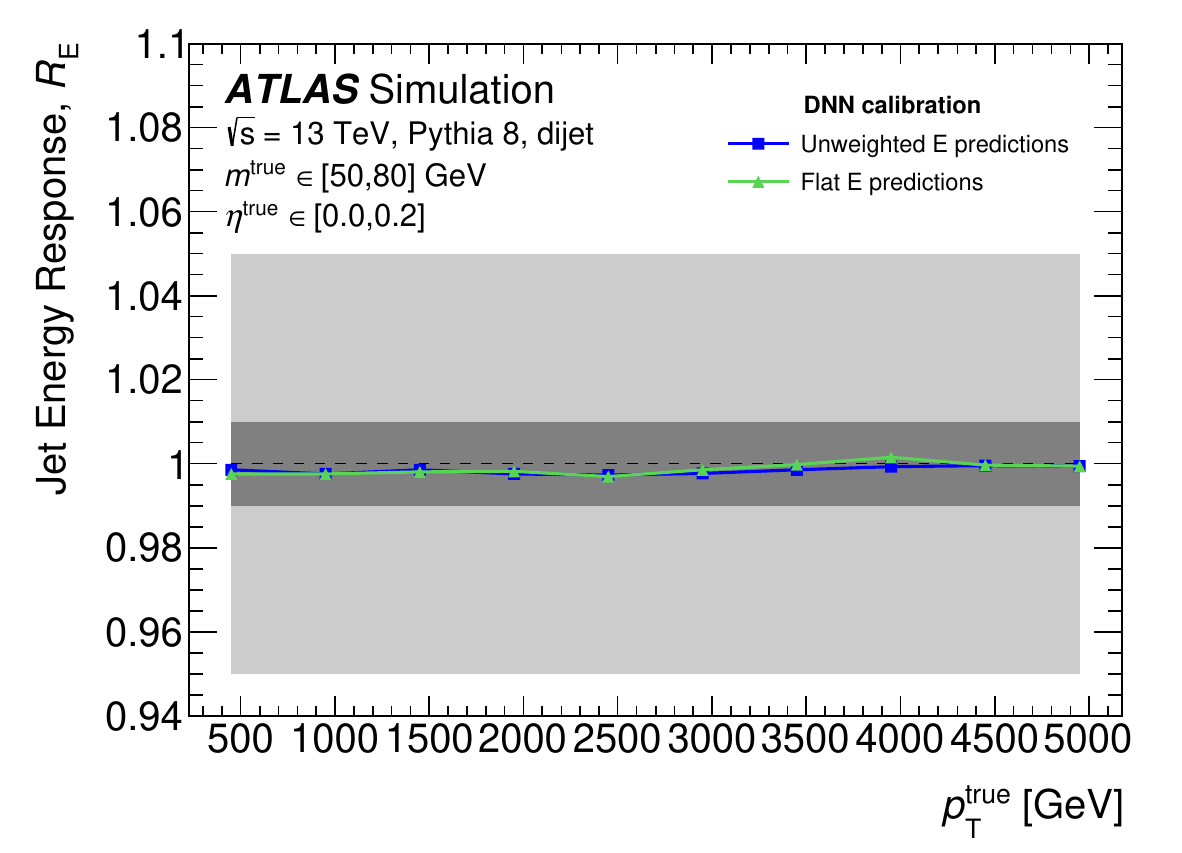} }}
\subfloat[\centering]{{\includegraphics[width=7.55cm]{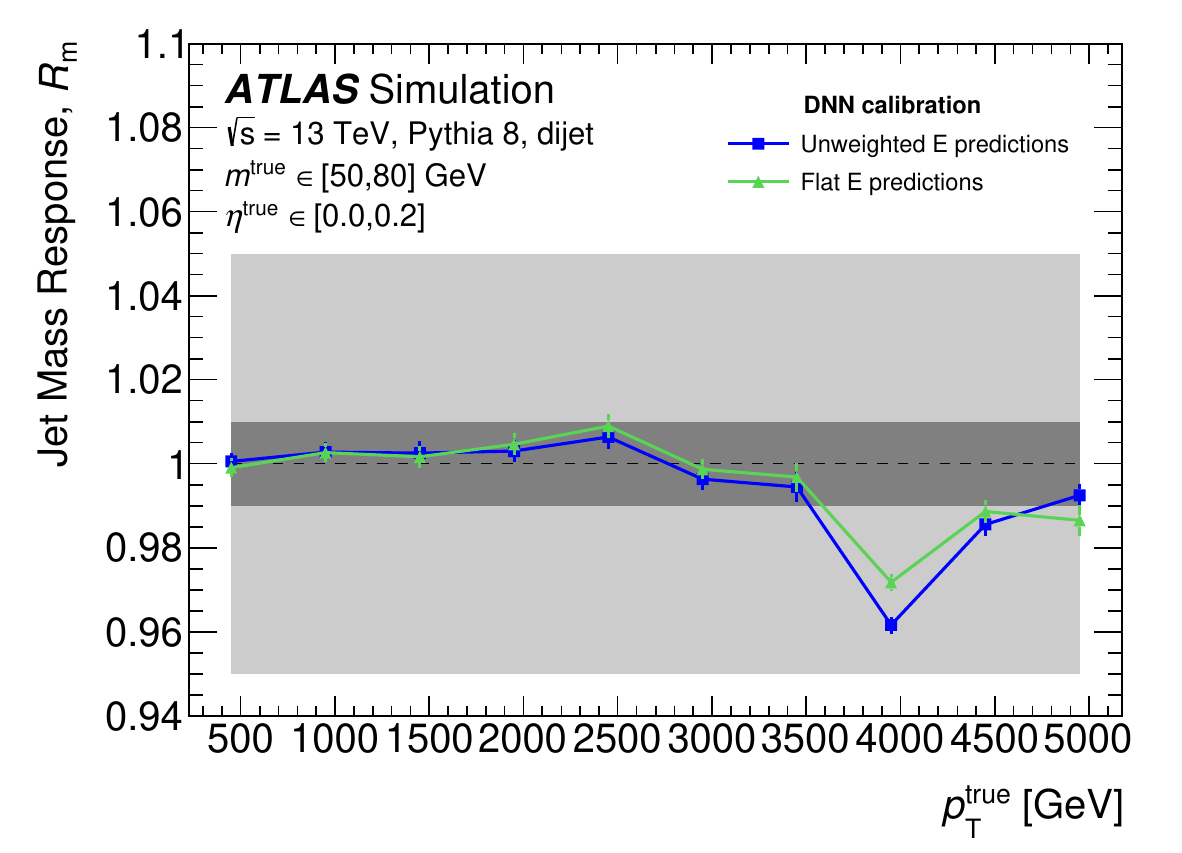} }}
\caption{Normalised distributions of (a) the jet energy before (black) and after (red) re-weighting. And the corresponding (b) jet energy and (c) mass responses obtained with the DNN calibration as a function of $\pT^\true$ for these two energy distributions, the unweighted distribution (squares) (as used for training) and the flat-energy weighted distribution (triangles). The horizontal bands represent 1$\%$ (dark) and 5$\%$ (light) up and down divergences from closure.}
\label{fig:emResp_LogEW_QCD}
\end{figure}

The DNN is trained and validated on a specific set of jets produced
through a particular Monte Carlo process with a large
range of energy.  This set of jets leads to specific distributions of
the input variables. The DNN calibration performance should be
independent of the distribution to which it is applied, so it can
be used in different physics analyses with different event selections
resulting in different input distributions. Since the calibration is
already assessed as a function of true \pT, the impact of the
distribution of this variable (or the energy one) is not expected to
be strong.  However, other input variables are correlated with the \pT
and are not binned in the performance tests; their distribution
shapes may thus impact the calibration.  To verify this, the
performance is re-evaluated by changing the response distributions
using statistical weights when constructing the response
histograms. The weights are chosen to obtain a flat jet energy
distribution, entirely different from the original one used in the
training as illustrated in
Figure~\ref{fig:emResp_LogEW_QCD_dist}.\footnote{Distributions are
shown after transformation and normalisation, e.g., the histogram shows
$f(E) = s \times \log(E) + o$ where $s$ and $o$ are such that $f(E)$ is mostly distributed in $[-1,1]$} The performance is then evaluated in
large \pT bins similar to those that physics analysis may use and inside
which distribution differences may have significant effects.
Figure~\ref{fig:emResp_LogEW_QCD} (b, c) shows the jet energy and
mass responses obtained with the DNN calibration with (square
markers) and without (triangular markers) re-weighting. The
differences between the two calibrated response closures are very
small, proving that the DNN calibration does not depend significantly on the input jet
distribution.
 
\paragraph{Pile-up dependency:}
 
\begin{figure}[htbp]
\centering
\subfloat[\centering]{{\includegraphics[width=7.55cm]{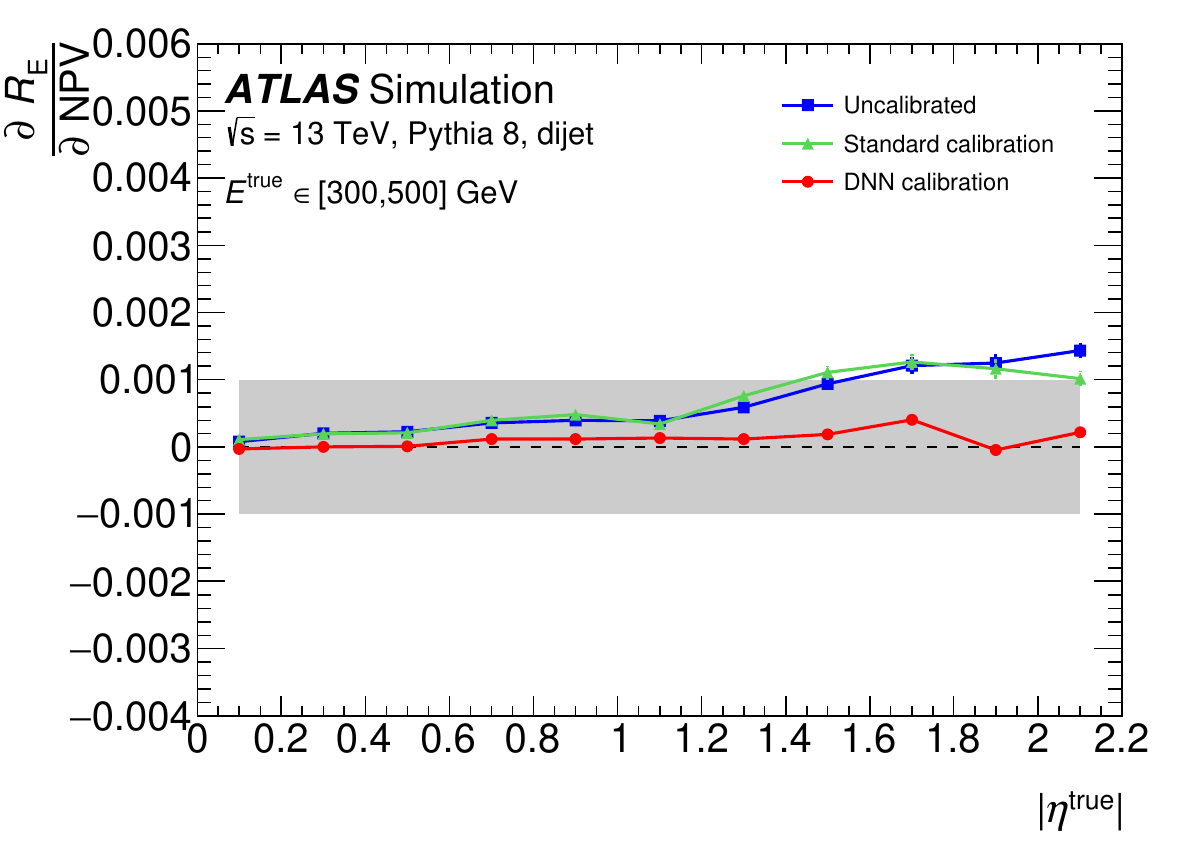} }}
\qquad
\subfloat[\centering]{{\includegraphics[width=7.55cm]{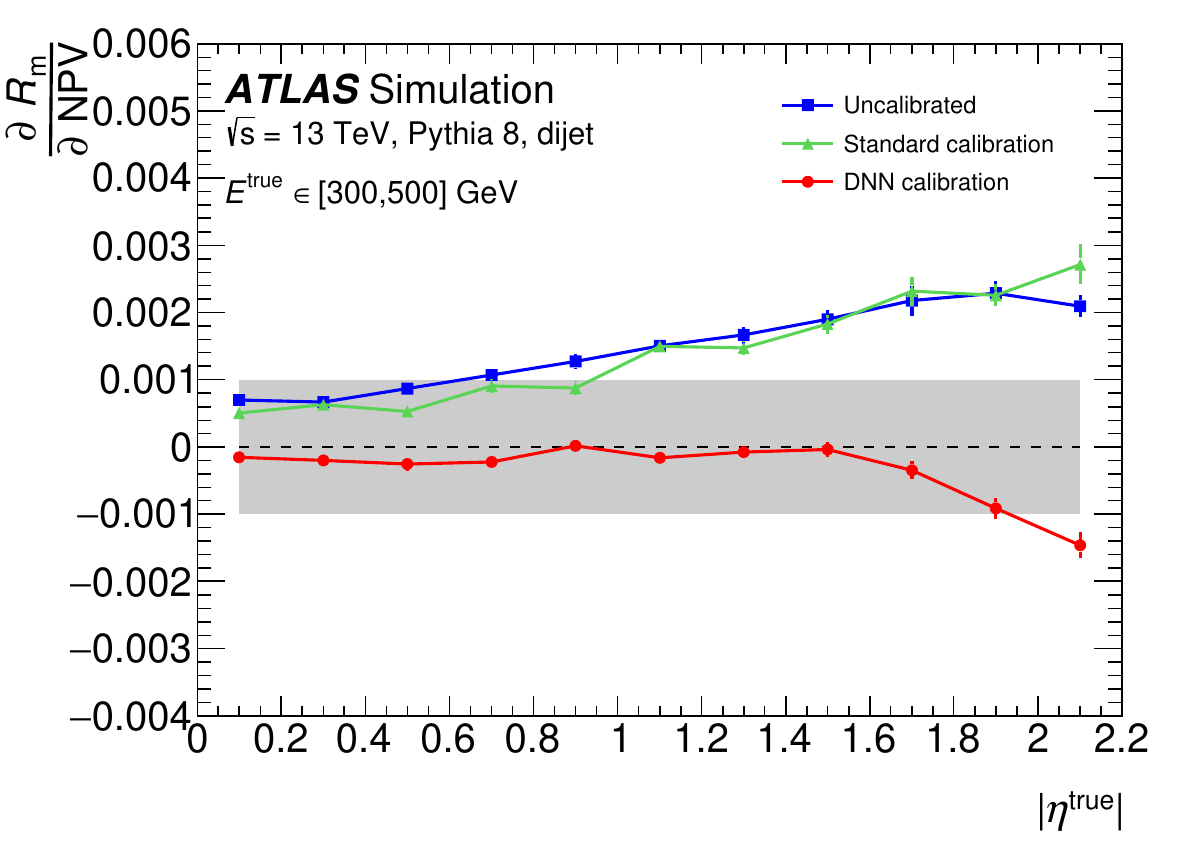} }}
\caption{The dependence of (a) the jet energy and (b)  the mass response on $\text{NPV}$ for different calibrations (squares: no calibration, triangles: standard calibration, circles: DNN predicted calibration) as a function of $|\eta^\true|$. The horizontal band represent $0.1\%$ up and down divergences from a null gradient.}
\label{fig:emRespGrad_NPV_QCD}
\end{figure}

The impact of the pile-up conditions is tested and evaluated by
considering the dependency of the response on two quantifying
variables: the number of reconstructed primary vertices in an event, $\text{NPV}$, and the average
number of collisions per bunch crossing, $\mu$. For this, the energy (or mass) response
$R_{E_\calib}$ is estimated in bins of $(v_\PU,E^\true,\eta^\true)$
where $v_\PU$ is NPV or $\mu$. A linear fit to $R_{E_\calib}$ against
$v_\PU$ is then performed for each $(E^\true,\eta^\true)$ and the
corresponding slope, the \emph{gradient} relative to $v_\PU$ denoted as
$\frac{\partial R_E}{\partial v_\PU}$, is extracted. Figure~\ref{fig:emRespGrad_NPV_QCD} presents
the gradient relative to $\text{NPV}$ of the energy and mass calibrated
response modes as a function of $\eta$. For both the energy
and the mass, the DNN calibration shows smaller gradients than the standard calibration. Similar results are obtained for $\mu$.
 
This shows that the grooming procedure performed during the jet reconstruction is not enough
to suppress pile-up effects, while the DNN is able to do so by
including $\text{NPV}$ and $\mu$ as input variables.
 
\paragraph{Jet flavour dependence:}
 
As discussed above, the DNN performance is assessed with
different topologies (QCD, $W/Z$, top-quark, or Higgs boson jets). It is also important
to understand the dependency of the DNN calibrations on the actual
parton that initiated the jet, also known as its flavour (light-quark ($u, d, s, c$),
$b$-quark or gluon jets), as different responses for different flavours will translate
into systematic uncertainties through the uncertainty in the flavour composition of the jets.
 
\begin{figure}[htbp]
\centering
\subfloat[\centering Standard calibration]{{\includegraphics[width=7.55cm]{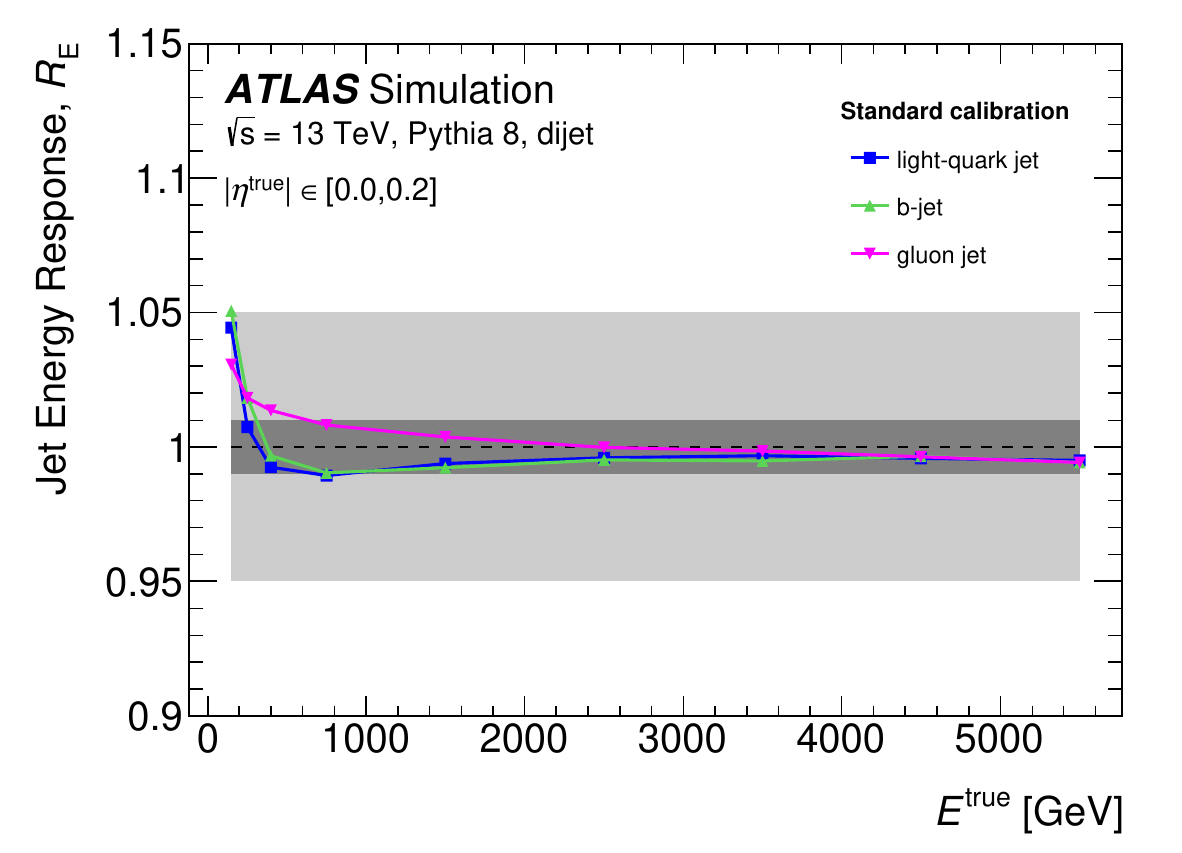} }}
\qquad
\subfloat[\centering Standard calibration]{{\includegraphics[width=7.55cm]{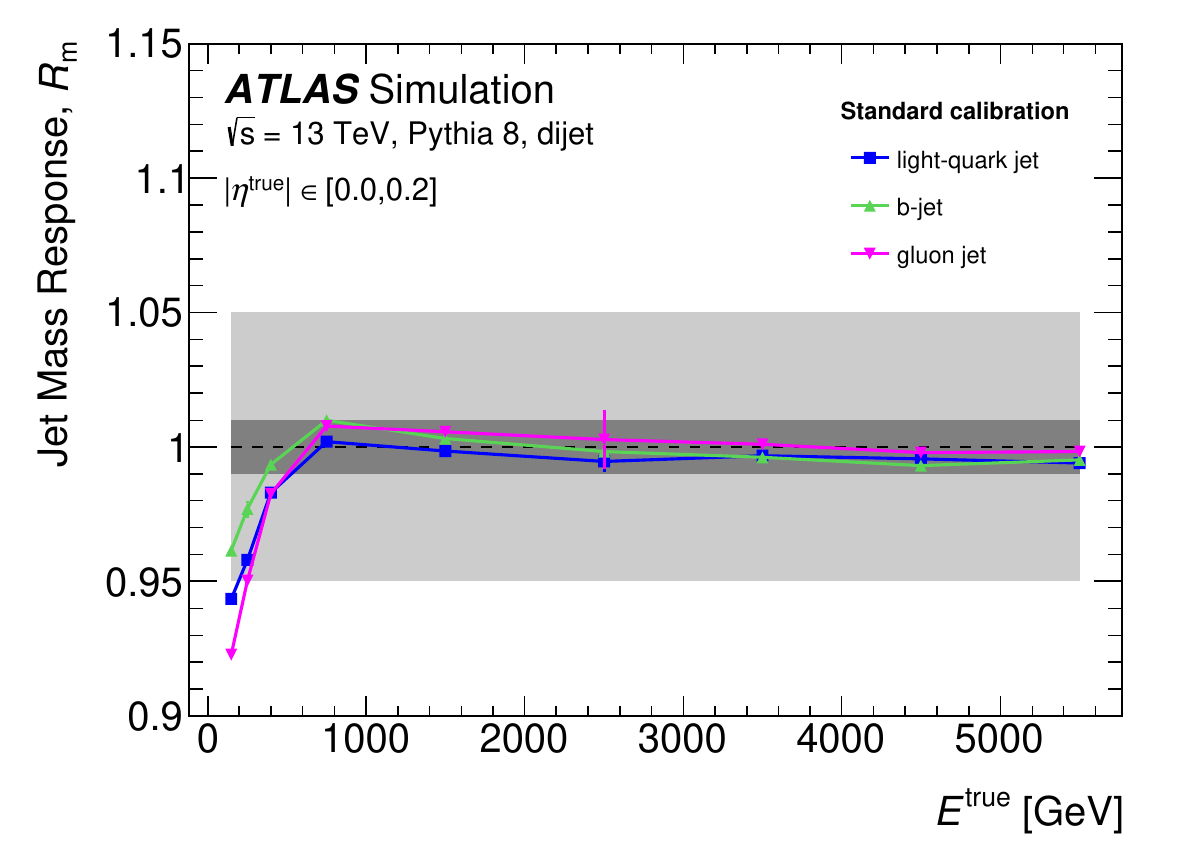} }}
\qquad
\subfloat[\centering DNN calibration]{{\includegraphics[width=7.55cm]{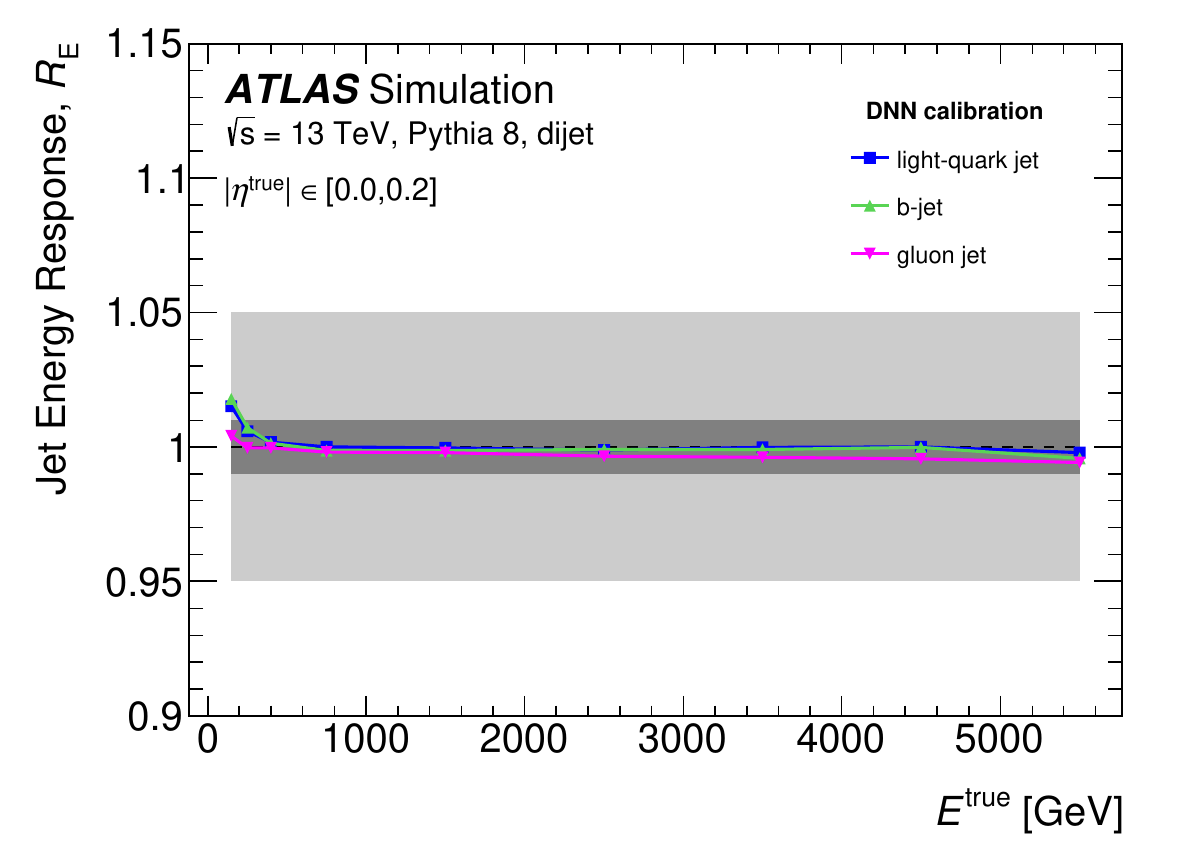} }}
\qquad
\subfloat[\centering DNN calibration]{{\includegraphics[width=7.55cm]{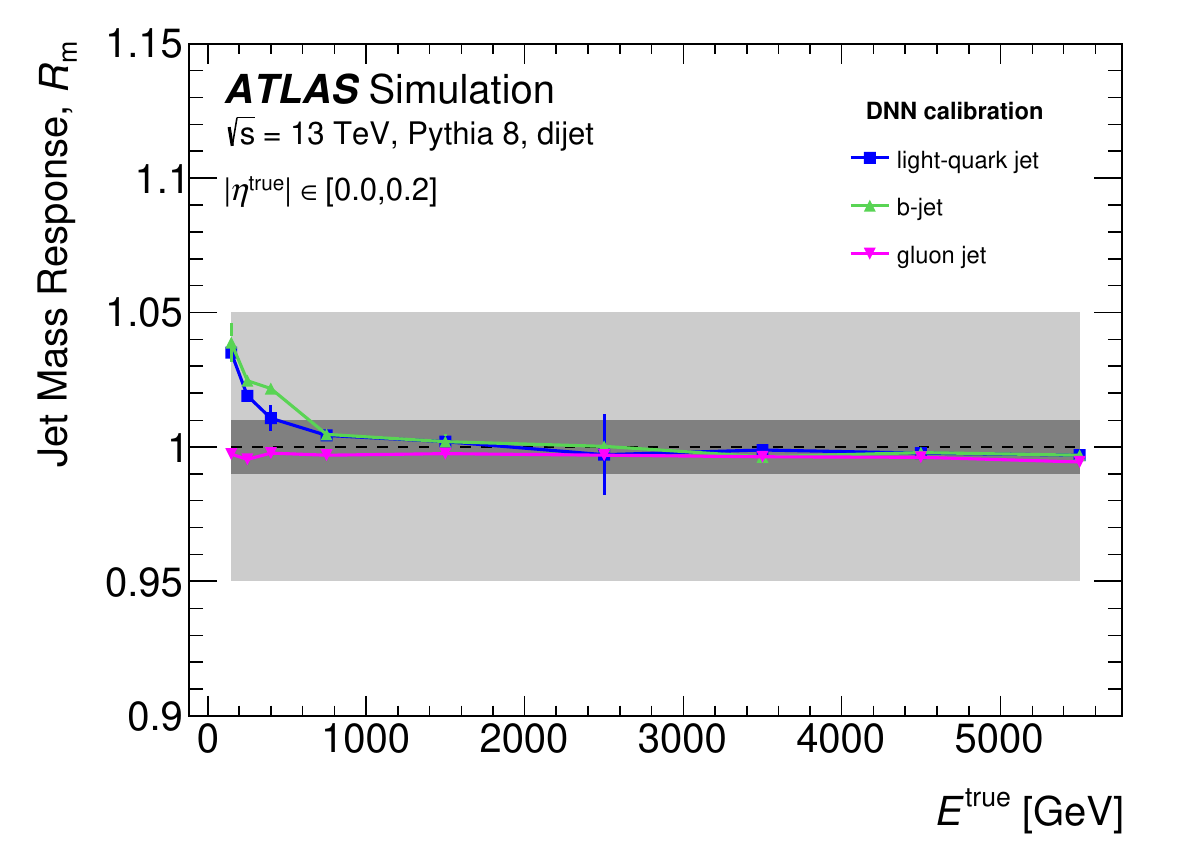} }}
\caption{Jet energy (left) and mass (right) responses versus $E^\true$ for different jet flavours, calibrated with the standard calibration (a, b) and the DNN calibration (c, d). The horizontal bands represent 1$\%$ (dark) and 5$\%$ (light) up and down divergences from closure.}
\label{fig:emResp_PID_QCD}
\end{figure}

Figure~\ref{fig:emResp_PID_QCD} presents the energy and mass calibrated responses as a function of the
jet energy for different jet flavours, comparing the DNN calibration
to the standard calibration. The jet flavour is defined here as the
flavour of the parton with the highest energy in the parton shower. These show that
the DNN predicts better calibrations for the different partons, with
smaller inter-flavour differences relative to the standard
calibration, especially at low \pT.
 
\paragraph{Dependency on the generator:}
The DNN calibration will be applied to all simulated samples used in
ATLAS analyses that are produced with a variety of generators, as
well as jets in real data events. Therefore it is desirable to evaluate
if the performance of the network is strongly affected by the modelling
of jets within \PYTHIA{8}.  Generators have specific ways of evaluating
the parton shower and hadronisation resulting in differences in the
predicted particle flow and simulated detector showers. Since such
differences can be representative of those between simulation and real
data, it is important to verify they are not detrimental to the
performance of the DNN calibration.
 
\begin{figure}[htbp]
\centering
\subfloat[\centering]{{\includegraphics[width=7.55cm]{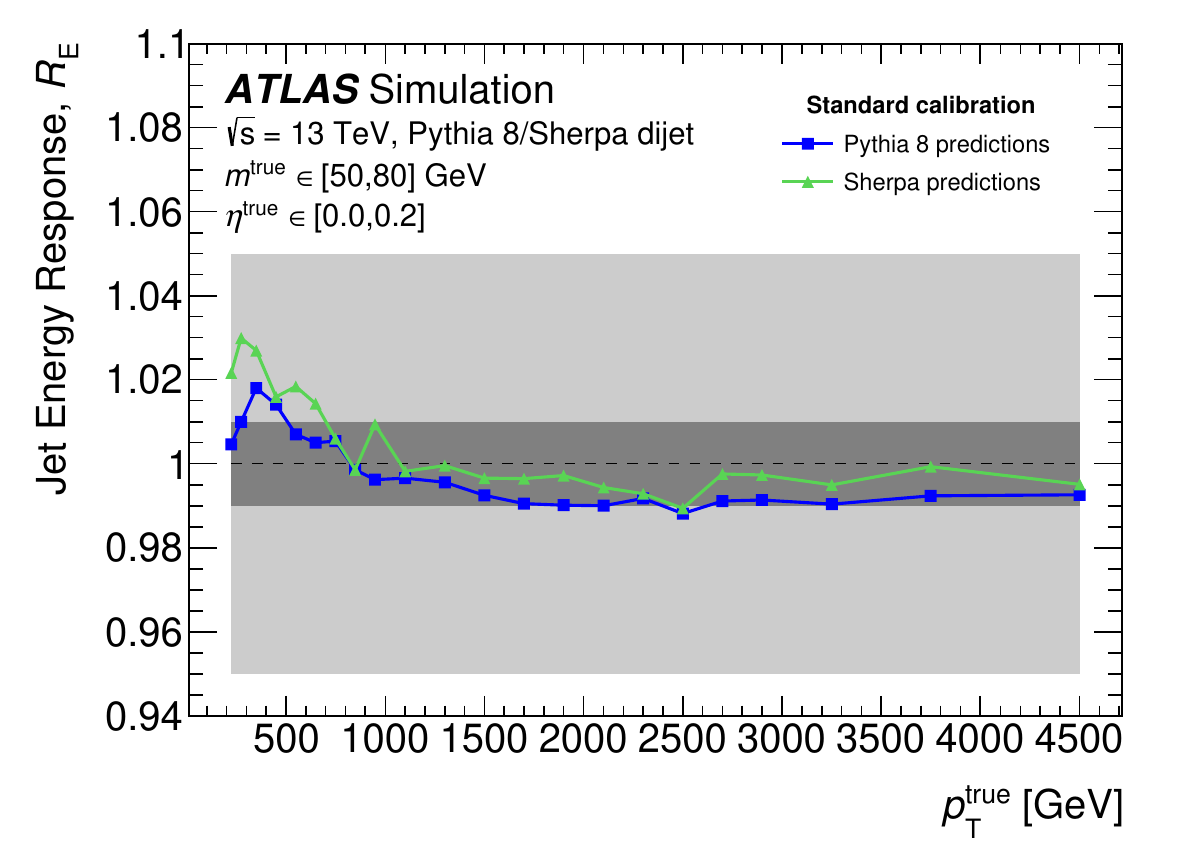} }}
\qquad
\subfloat[\centering]{{\includegraphics[width=7.55cm]{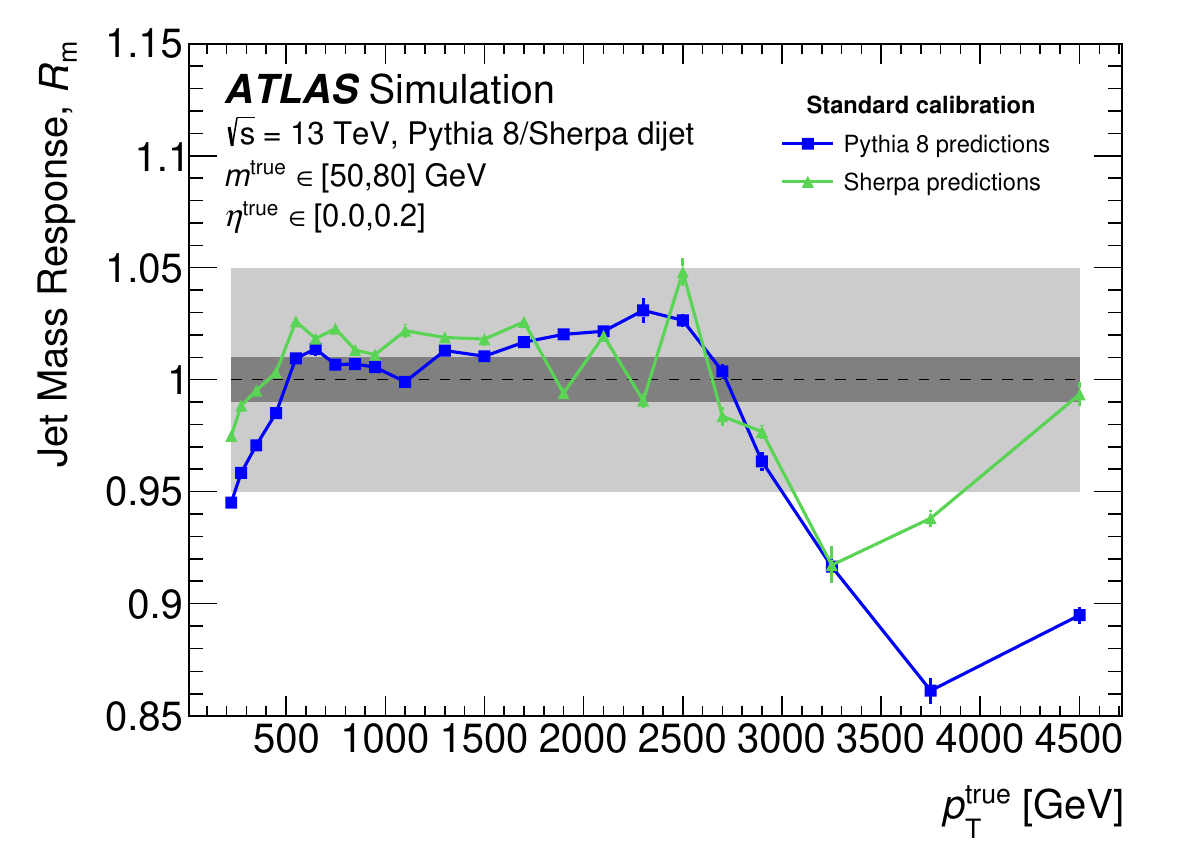} }}
\qquad
\subfloat[\centering]{{\includegraphics[width=7.55cm]{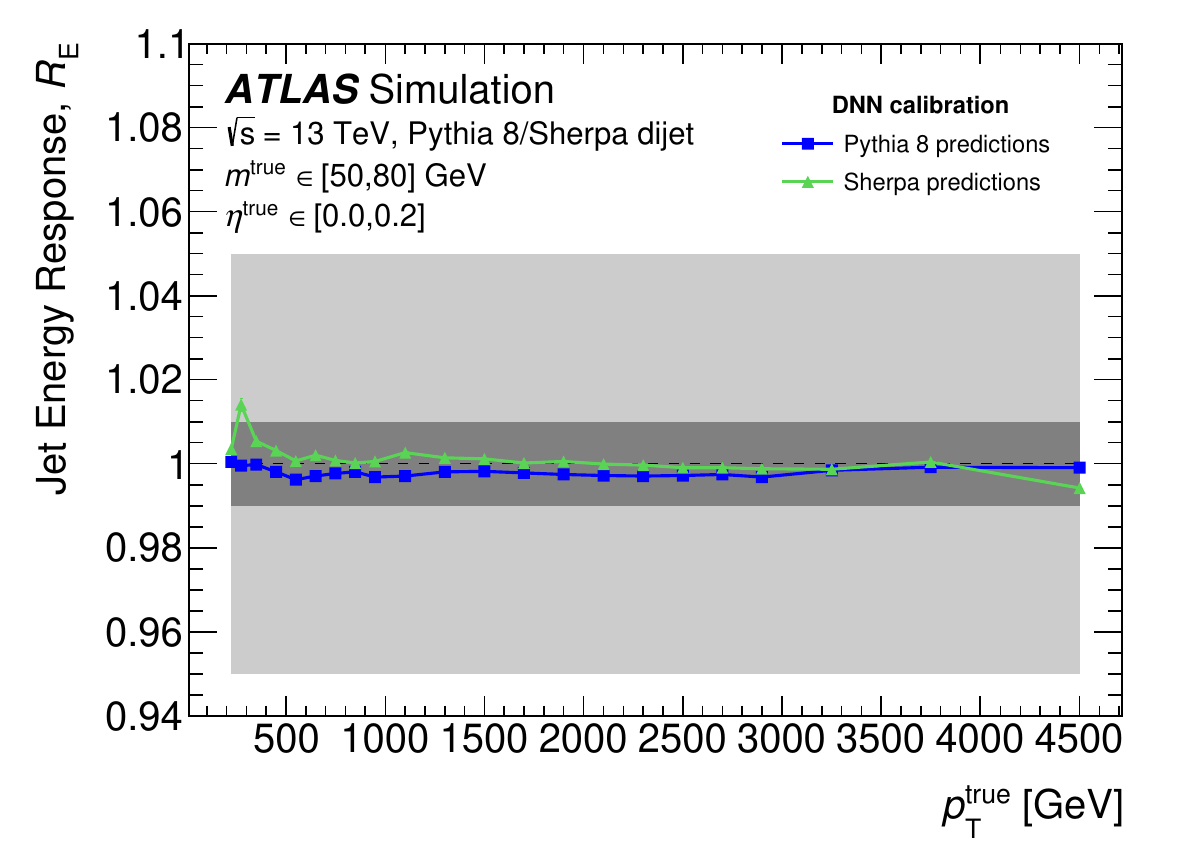} }}
\qquad
\subfloat[\centering]{{\includegraphics[width=7.55cm]{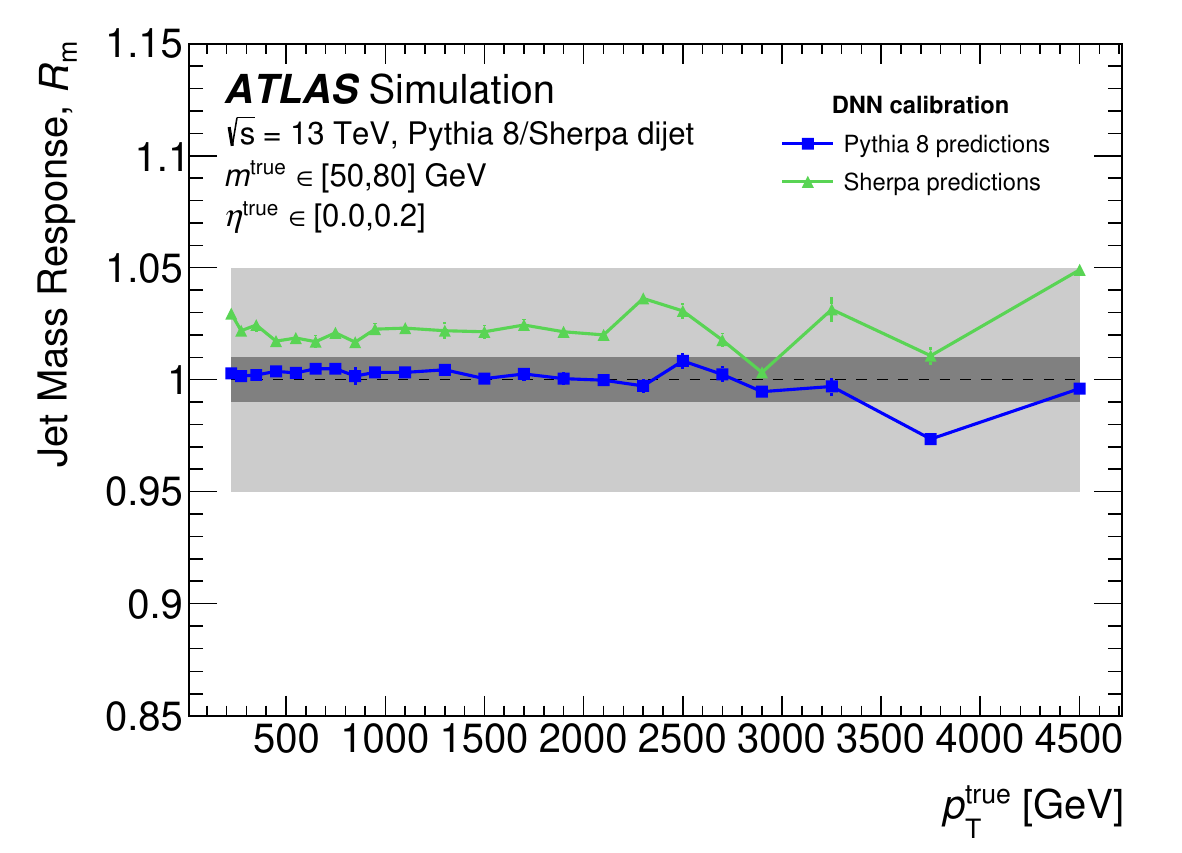} }}
\caption{Jet energy and mass responses versus $\pT^\true$ calibrated with (a, b) the standard calibration and (c, d) with the DNN predictions for a dijet data sample generated with \PYTHIA (squares) or \SHERPA (triangles). The horizontal bands represent 1$\%$ (dark) and 5$\%$ (light) up and down divergences from closure.}
\label{fig:emRespvspT_QCD_Sherpa}
\end{figure}

A data sample of \SHERPA-generated
dijets with the same data preparation as described in Section~\ref{subsec:dataprep} is
used to compute jet energy and mass responses with the standard
calibration and the DNN calibration.\footnote{The DNN is trained with the \PYTHIA sample as discussed in the previous sections.} Both are then compared with the performance obtained with the \PYTHIA
dijet data sample. Figure~\ref{fig:emRespvspT_QCD_Sherpa} shows this comparison. The DNN energy calibration
shows almost identical performance for the two generators, except at low $\pT^\true$ where a small under-performance
is observed with the \SHERPA data sample although closure is however still obtained. Moreover, the DNN
calibration shows a lesser dependency on the generator than the standard calibration for
the energy. For the mass calibration with the \SHERPA
data sample, the DNN response closes within 5\%  but outside 1\%
in all the $\pT^\true$ range. This can be expected  as the jet's
mass is more dependent on its particle flow structure. This
under-performance is observed to be reduced in higher mass bins. The standard mass calibration
shows smaller differences between the two generators but overall, the
DNN mass calibration is still better than the standard one for both the \PYTHIA and \SHERPA generators.
 
\section{Limitations and possible solutions}
\label{sec:prospects}
Despite the satisfying results obtained with the calibration methodology
presented above, some limitations and open questions are
identified. They are discussed together with possible solutions in this section.
 
\subsection{Training convergence and ending criterion}
 
A weakness in this DNN training procedure is the lack of a clear criterion
to mark the final convergence of the network and end the
training.
It is observed that early interruptions of the training (as soon as
the losses start to fluctuate around a minimum) result in worse
calibration performance in terms of JES, JMS or resolution in
significant parts of the phase space. Instead, training for numerous
additional epochs (such as described in
Section~\ref{subsec:training}) during which the training loss remains
essentially constant is found to be necessary.
A possible explanation is
that the goodness of the performance is based on relatively strict
expectations (e.g., closure within 1\% for the energy response)
to be obtained over all the ($E$, $\eta$, $m$) phase space while only a
part of it is statistically dominant in the training sample. As a
consequence, the global loss can be mostly determined by this dominant
phase space region in which the optimum is quickly reached, while it
requires more training to improve the sparsely populated regions. This
behaviour also resembles the `grokking phenomenon' observed in
some DNN problems where sometimes performance suddenly increases after a
very long training period~\cite{nanda2023progress}.
 
Although the present study shows that training long enough allows
coverage of the full phase space, a robust methodology would require a well-understood
criterion to stop the training. Evaluating the calibration performance across all the phase
space is not a practical option as it is too
costly to perform automatically at each epoch. An interesting approach
would be to evaluate the loss in predefined, well-chosen bins (in
terms of statistical contents and of physics relevance) of the phase
space. Monitoring the losses in these bins with training and
validation samples would possibly be more indicative than monitoring the
virtually flat total loss.
 
\subsection{Calibration of heavy boosted jets}
 
During the validation of the DNN calibration performance for $W/Z$, Higgs boson and top-quark jets,
non-closures of up to a few percent are observed for the energy (see Figures~\ref{fig:emResp_boosted}). While this under-performance is also visible for the standard calibration, the DNN calibration performance is slightly worse in these cases. Contrary to the DNN calibration presented here, the standard calibration is obtained using different types of jets. At high \pT (above $\sim 500$~\GeV) QCD jets are used, while at low \pT the calibration is only derived from $W/Z$ jets as this choice leads to better calibration performance for both $W/Z$ jets and QCD jets.
 
A similar approach with mixed training samples is possible with the
DNN\@. However, simply adding $W/Z$ samples to the training does not
improve -- or even degrades -- the performance for mass regions away
from the $W$ and $Z$ masses. This issue could be mitigated with the
availability of special pseudo-$W/Z$ samples having a flat mass
distribution. Alternatively, an ad hoc solution mimicking the standard
calibration is possible by adding a subnetwork to the already
QCD-trained DNN\@. This subnetwork would branch from the input layer with $\eta$ annotation, produce output
values used as multiplicative correction factors to the main
network predictions and enforce these factors to be 1.0 for large \pT jets
(typically above $500$~\GeV) to maintain the good performance of the DNN in
this phase space. The subnetwork could then be trained on $W/Z$ samples, while keeping the main
network frozen to improve the lower \pT regions.
 
\subsection{Importance of the features}
\begin{figure}[htbp]
\begin{center}
\includegraphics[width=16cm,keepaspectratio]{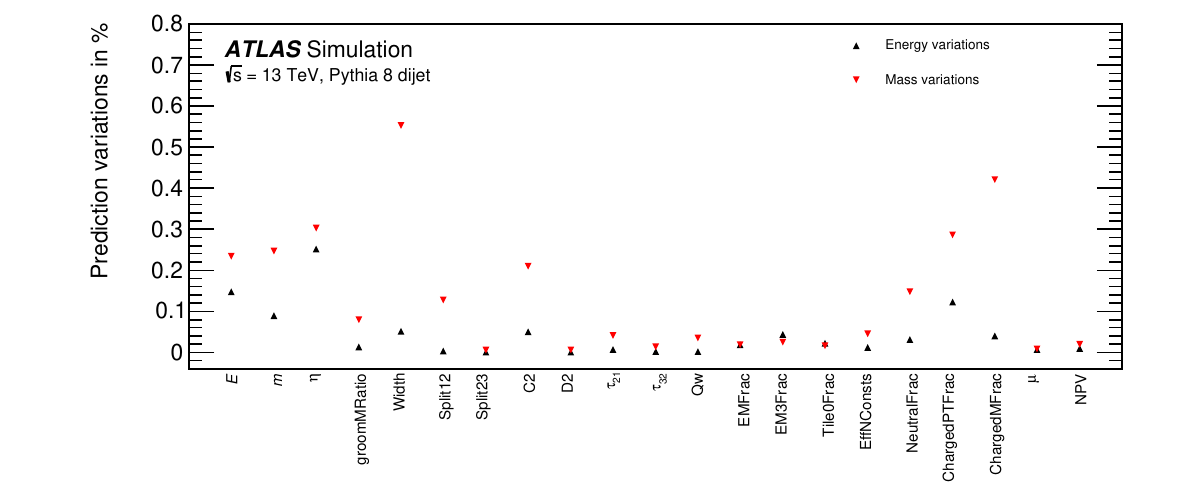}
\end{center}
\caption{The variation of the energy (upward triangles) and mass (downward triangles)
predictions when input features are varied (see the text for details).}
\label{fig:feature_importance}
\end{figure}

Insights into which input features are the most
influential on the DNN predictions are important for two
reasons. Firstly, they generally help to understand the jet energy
and mass responses.
Secondly, they can indicate if some features could
be removed from the inputs, potentially leading to a simpler DNN or even to better performance if
the removed variables are confusing the network.
A simple way to quantify the importance of a feature is to measure the
average of the relative variation in the prediction when this feature is
systematically varied for each input jet.
A ranking of the different features obtained by following this method is shown
in Figure~\ref{fig:feature_importance} where each feature is
varied by $\pm 1\%$ of the standard deviation of its overall distribution.
Both the energy and mass prediction variations are displayed
for each feature.
As expected, the variables $E$, $\eta$, and $m$ have a large
influence on the predictions according to this measure.
Some substructure features\footnote{The Width, groomMRatio,
ChargedPTFrac, ChargedMFrac, Split12, and C2.} have a stronger impact on
the mass prediction than on the energy prediction. This is also expected
since the mass depends on the structure of the particle flow objects in the jet.
While by this metric $\mu$ and $\text{NPV}$ do not seem so important,
it was demonstrated in Section~\ref{subsec:validation} that these features are
important as they allow the DNN calibration to remove the response dependence on these variables.
Further investigations are therefore necessary to understand more precisely
the importance of the features. Those could
consist of considering the magnitude of the DNN gradient with respect
to its inputs in bins of \pT or mass, or to apply the SHAP methodology.\footnote{SHapley Additive exPlanations~\cite{SHAP}.}


\section{Conclusion}
 
A full implementation of a deep neural network based simultaneous calibration of the
energy and mass of \largeR jets with the ATLAS detector is
presented. It solves problems that are specific to the calibration
task. Firstly, rather than segmenting the calibration in angular slices,
a single network covers multiple detector regions and a dedicated
encoding of the jet angular direction is included to help it to adapt
to the corresponding strongly varying response. Secondly, the network
architecture is designed to efficiently accommodate both the energy
and the mass predictions. Finally, loss functions are carefully chosen
so that the DNN prediction converges as desired toward the most
probable value of the energy or mass response distribution; this also
implies non-trivial training procedures involving variants of the loss
functions.
 
The resulting calibration is compared to the standard calibration on
QCD jets and on other jet topologies not seen during the DNN training.
Over all the phase space and for all processes of interest, the DNN
is found to achieve better energy and mass scale closure and similar
or better resolution, with a typical 30\% improvement in the energy
resolution for $p_T>\qty{500}{\GeV}$. The DNN calibration is also shown
to be robust against various tests such as pile-up dependency or
changes to the sample distribution. Finally, some limitations and possible improvements
are discussed. Overall, as this simulation-based method can directly
replace the current analogous steps of the ATLAS \largeR jet
calibration chain, no adaptation of the data-based methodology is required
and this DNN procedure can be used for the calibration of
large-radius jets used in future ATLAS analyses.


\section*{Acknowledgements}


We thank CERN for the very successful operation of the LHC and its injectors, as well as the support staff at
CERN and at our institutions worldwide without whom ATLAS could not be operated efficiently.
 
The crucial computing support from all WLCG partners is acknowledged gratefully, in particular from CERN, the ATLAS Tier-1 facilities at TRIUMF/SFU (Canada), NDGF (Denmark, Norway, Sweden), CC-IN2P3 (France), KIT/GridKA (Germany), INFN-CNAF (Italy), NL-T1 (Netherlands), PIC (Spain), RAL (UK) and BNL (USA), the Tier-2 facilities worldwide and large non-WLCG resource providers. Major contributors of computing resources are listed in Ref.~\cite{ATL-SOFT-PUB-2023-001}.
 
We gratefully acknowledge the support of ANPCyT, Argentina; YerPhI, Armenia; ARC, Australia; BMWFW and FWF, Austria; ANAS, Azerbaijan; CNPq and FAPESP, Brazil; NSERC, NRC and CFI, Canada; CERN; ANID, Chile; CAS, MOST and NSFC, China; Minciencias, Colombia; MEYS CR, Czech Republic; DNRF and DNSRC, Denmark; IN2P3-CNRS and CEA-DRF/IRFU, France; SRNSFG, Georgia; BMBF, HGF and MPG, Germany; GSRI, Greece; RGC and Hong Kong SAR, China; ISF and Benoziyo Center, Israel; INFN, Italy; MEXT and JSPS, Japan; CNRST, Morocco; NWO, Netherlands; RCN, Norway; MEiN, Poland; FCT, Portugal; MNE/IFA, Romania; MESTD, Serbia; MSSR, Slovakia; ARRS and MIZ\v{S}, Slovenia; DSI/NRF, South Africa; MICINN, Spain; SRC and Wallenberg Foundation, Sweden; SERI, SNSF and Cantons of Bern and Geneva, Switzerland; MOST, Taipei; TENMAK, T\"urkiye; STFC, United Kingdom; DOE and NSF, United States of America.
 
Individual groups and members have received support from BCKDF, CANARIE, CRC and DRAC, Canada; CERN-CZ, PRIMUS 21/SCI/017 and UNCE SCI/013, Czech Republic; COST, ERC, ERDF, Horizon 2020, ICSC-NextGenerationEU and Marie Sk{\l}odowska-Curie Actions, European Union; Investissements d'Avenir Labex, Investissements d'Avenir Idex and ANR, France; DFG and AvH Foundation, Germany; Herakleitos, Thales and Aristeia programmes co-financed by EU-ESF and the Greek NSRF, Greece; BSF-NSF and MINERVA, Israel; Norwegian Financial Mechanism 2014-2021, Norway; NCN and NAWA, Poland; La Caixa Banking Foundation, CERCA Programme Generalitat de Catalunya and PROMETEO and GenT Programmes Generalitat Valenciana, Spain; G\"{o}ran Gustafssons Stiftelse, Sweden; The Royal Society and Leverhulme Trust, United Kingdom.
 
In addition, individual members wish to acknowledge support from CERN: European Organization for Nuclear Research (CERN PJAS); Chile: Agencia Nacional de Investigación y Desarrollo (FONDECYT 1190886, FONDECYT 1210400, FONDECYT 1230987); China: National Natural Science Foundation of China (NSFC - 12175119, NSFC 12275265); European Union: European Research Council (ERC - 948254, ERC 101089007), Horizon 2020 Framework Programme (MUCCA - CHIST-ERA-19-XAI-00), Italian Center for High Performance Computing, Big Data and Quantum Computing (ICSC, NextGenerationEU); France: Agence Nationale de la Recherche (ANR-20-CE31-0013, ANR-21-CE31-0013, ANR-21-CE31-0022), Investissements d'Avenir Labex (ANR-11-LABX-0012); Germany: Baden-Württemberg Stiftung (BW Stiftung-Postdoc Eliteprogramme), Deutsche Forschungsgemeinschaft (DFG - 469666862, DFG - CR 312/5-1); Italy: Istituto Nazionale di Fisica Nucleare (ICSC, NextGenerationEU); Japan: Japan Society for the Promotion of Science (JSPS KAKENHI 22H01227, JSPS KAKENHI 22KK0227, JSPS KAKENHI JP21H05085, JSPS KAKENHI JP22H04944); Netherlands: Netherlands Organisation for Scientific Research (NWO Veni 2020 - VI.Veni.202.179); Norway: Research Council of Norway (RCN-314472); Poland: Polish National Agency for Academic Exchange (PPN/PPO/2020/1/00002/U/00001), Polish National Science Centre (NCN 2021/42/E/ST2/00350, NCN OPUS nr 2022/47/B/ST2/03059, NCN UMO-2019/34/E/ST2/00393, UMO-2020/37/B/ST2/01043, UMO-2021/40/C/ST2/00187, UMO-2022/47/O/ST2/00148); Slovenia: Slovenian Research Agency (ARIS grant J1-3010); Spain: BBVA Foundation (LEO22-1-603), Generalitat Valenciana (Artemisa, FEDER, IDIFEDER/2018/048), La Caixa Banking Foundation (LCF/BQ/PI20/11760025), Ministry of Science and Innovation (RYC2019-028510-I, RYC2020-030254-I), PROMETEO and GenT Programmes Generalitat Valenciana (CIDEGENT/2019/023, CIDEGENT/2019/027); Sweden: Swedish Research Council (VR 2022-03845, VR 2022-04683), Knut and Alice Wallenberg Foundation (KAW 2017.0100, KAW 2018.0157, KAW 2019.0447); Switzerland: Swiss National Science Foundation (SNSF - PCEFP2\_194658); United Kingdom: Leverhulme Trust (Leverhulme Trust RPG-2020-004); United States of America: Neubauer Family Foundation.


\printbibliography
 
\clearpage
 
\begin{flushleft}
\hypersetup{urlcolor=black}
{\Large The ATLAS Collaboration}

\bigskip

\AtlasOrcid[0000-0002-6665-4934]{G.~Aad}$^\textrm{\scriptsize 102}$,
\AtlasOrcid[0000-0001-7616-1554]{E.~Aakvaag}$^\textrm{\scriptsize 16}$,
\AtlasOrcid[0000-0002-5888-2734]{B.~Abbott}$^\textrm{\scriptsize 120}$,
\AtlasOrcid[0000-0002-1002-1652]{K.~Abeling}$^\textrm{\scriptsize 55}$,
\AtlasOrcid[0000-0001-5763-2760]{N.J.~Abicht}$^\textrm{\scriptsize 49}$,
\AtlasOrcid[0000-0002-8496-9294]{S.H.~Abidi}$^\textrm{\scriptsize 29}$,
\AtlasOrcid[0000-0002-9987-2292]{A.~Aboulhorma}$^\textrm{\scriptsize 35e}$,
\AtlasOrcid[0000-0001-5329-6640]{H.~Abramowicz}$^\textrm{\scriptsize 151}$,
\AtlasOrcid[0000-0002-1599-2896]{H.~Abreu}$^\textrm{\scriptsize 150}$,
\AtlasOrcid[0000-0003-0403-3697]{Y.~Abulaiti}$^\textrm{\scriptsize 117}$
% ...
\bigskip
\\

% ...
$^{16}$Department for Physics and Technology, University of Bergen, Bergen; Norway.\\
% ...
$^{29}$Physics Department, Brookhaven National Laboratory, Upton NY; United States of America.\\
% ...
$^{35}$$^{(a)}$Facult\'e des Sciences Ain Chock, R\'eseau Universitaire de Physique des Hautes Energies - Universit\'e Hassan II, Casablanca;$^{(b)}$Facult\'{e} des Sciences, Universit\'{e} Ibn-Tofail, K\'{e}nitra;$^{(c)}$Facult\'e des Sciences Semlalia, Universit\'e Cadi Ayyad, LPHEA-Marrakech;$^{(d)}$LPMR, Facult\'e des Sciences, Universit\'e Mohamed Premier, Oujda;$^{(e)}$Facult\'e des sciences, Universit\'e Mohammed V, Rabat;$^{(f)}$Institute of Applied Physics, Mohammed VI Polytechnic University, Ben Guerir; Morocco.\\
% ...
$^{49}$Fakult\"{a}t Physik , Technische Universit{\"a}t Dortmund, Dortmund; Germany.\\
% ...
$^{55}$II. Physikalisches Institut, Georg-August-Universit\"{a}t G\"ottingen, G\"ottingen; Germany.\\
% ...
$^{102}$CPPM, Aix-Marseille Universit\'e, CNRS/IN2P3, Marseille; France.\\
% ...
$^{117}$Department of Physics, New York University, New York NY; United States of America.\\
% ...
$^{120}$Homer L. Dodge Department of Physics and Astronomy, University of Oklahoma, Norman OK; United States of America.\\
% ...
$^{150}$Department of Physics, Technion, Israel Institute of Technology, Haifa; Israel.\\
$^{151}$Raymond and Beverly Sackler School of Physics and Astronomy, Tel Aviv University, Tel Aviv; Israel.\\
% ...
$^{a}$ Also Affiliated with an institute covered by a cooperation agreement with CERN.\\
$^{b}$ Also at An-Najah National University, Nablus; Palestine.\\
$^{c}$ Also at Borough of Manhattan Community College, City University of New York, New York NY; United States of America.\\
$^{d}$ Also at Center for High Energy Physics, Peking University; China.\\
$^{e}$ Also at Center for Interdisciplinary Research and Innovation (CIRI-AUTH), Thessaloniki; Greece.\\
$^{f}$ Also at Centro Studi e Ricerche Enrico Fermi; Italy.\\
% ...

\end{flushleft}


\end{document}